\documentclass[12pt,draftcls,onecolumn]{IEEEtran}

\renewcommand{\P}{\mbox{P}}
\renewcommand{\P}{{\mathbb{P}}}

\newcommand{\beq}{\begin{equation}}
\newcommand{\eeq}{\end{equation}}
\newcommand{\beqa}{\begin{eqnarray}}
\newcommand{\eeqa}{\end{eqnarray}}
\newcommand{\dfz}{\triangleq}
\newcommand{\bd}{\bm{d}}
\newcommand{\bx}{\bm{x}}
\newcommand{\by}{\bm{y}}
\newcommand{\bw}{\bm{w}}

\newcommand{\bv}{\bm{v}}

\newcommand{\VAR}{\textnormal{VAR}}

\usepackage{cite}
\usepackage{amsmath} 
\usepackage{amssymb} 
\usepackage{dsfont}

\usepackage{graphicx}
\usepackage{graphics}
\usepackage{bm}
\input{epsf}

\usepackage{color}

\newcommand{\E}{{\mathbb{E}}}

\usepackage{mathrsfs}

\begin{document}
%
\title{Distributed Detection over Adaptive Networks: Refined Asymptotics \\ and the Role of Connectivity}

%
\author{Vincenzo~Matta, Paolo~Braca, Stefano~Marano, Ali~H.~Sayed
\thanks{
V.~Matta and S.~Marano are with DIEM, University of Salerno,
via Giovanni Paolo II 132, I-84084, Fisciano (SA), Italy (e-mail: vmatta@unisa.it; marano@unisa.it).

P.~Braca is with NATO STO Centre for Maritime Research and Experimentation, La Spezia, Italy (e-mail: paolo.braca@cmre.nato.int).

A.~H.~Sayed is with the Electrical Engineering Department, University of California, Los Angeles, CA 90095 USA (e-mail: sayed@ee.ucla.edu).
}
\thanks{A short version of this work appeared in the conference publication~\cite{MattaBracaMaranoSayedICASSP2015}.}
}

\maketitle

\begin{abstract}
We consider distributed detection problems over adaptive networks, where dispersed agents learn continually from streaming data by means of local interactions. The requirement of adaptation allows the network of detectors to track drifts in the underlying hypothesis. The requirement of cooperation allows each agent to deliver a performance superior to what would be obtained if it were acting individually. The simultaneous requirements of adaptation and cooperation are achieved by employing diffusion algorithms with constant step-size $\mu$.
In~\cite{AdaptiveDetectionArxiv,ADD_ICASSP2014} some main features of adaptive distributed detection were revealed. By resorting to large deviations analysis, it was established that the Type-I and Type-II error probabilities of all agents vanish exponentially as functions of $1/\mu$, and that all agents share the same Type-I and Type-II error exponents. However, numerical evidences presented in~\cite{AdaptiveDetectionArxiv,ADD_ICASSP2014} showed that the theory of large deviations does not capture the fundamental impact of network connectivity on performance, and that additional tools and efforts are required to obtain accurate predictions for the error probabilities. This work addresses these open issues and extends the results of~\cite{AdaptiveDetectionArxiv,ADD_ICASSP2014} in several directions. 
By conducting a refined asymptotic analysis based on the mathematical framework of exact asymptotics, we arrive at a revealing and powerful understanding of the {\em universal behavior of distributed detection over adaptive networks}: as functions of $1/\mu$, the error (log-)probability curves corresponding to different agents stay nearly-parallel to each other (as already discovered in~\cite{AdaptiveDetectionArxiv,ADD_ICASSP2014}), however, these curves are ordered following a criterion reflecting the degree of connectivity of each agent. Depending on the combination weights, the more connected an agent is, the lower its error probability curve will be. 
The analysis provides explicit analytical formulas for the detection error probabilities and these expressions are also verified by means of extensive simulations.
We further enlarge the reference setting from the case of doubly-stochastic combination matrices considered in~\cite{AdaptiveDetectionArxiv,ADD_ICASSP2014}, to the more general and demanding setting of right-stochastic combination matrices; this extension poses new and interesting questions in terms of the interplay between the network topology, the combination weights, and the inference performance. 
The potential of the proposed methods is illustrated by application of the results to canonical detection problems, to typical network topologies, for both doubly-stochastic and right-stochastic combination matrices. Interesting and somehow unexpected behaviors emerge, and the lesson learned is that {\em connectivity matters}. 
\end{abstract}


\section{Introduction and Motivation}
\IEEEPARstart{A}{daptive} networks consist of spatially dispersed agents, learning continually from streaming data by means of distributed processing algorithms.
With a proper form of cooperation, each agent is able to deliver an inference performance superior to what would be obtained if it were acting individually.
When called to operate in complex and dynamical scenarios (e.g., in the presence of drifts in the statistical conditions, in the environmental conditions, and in the network topology, among other possibilities), the agents must be further endowed with strong adaptive capabilities, in order to respond in real-time to these variations. 

Several useful distributed inference solutions are available that meet these requirements. Particularly relevant to our setting are the diffusion implementations based on adaptive algorithms with constant step-size~\cite{LopesSayed,CattivelliSayedEstimation,SayedSPmag,SayedProcIEEE}. The use of a constant step-size (as opposed to the decaying step-size employed in the case of consensus algorithms~\cite{running-cons,asymptotic-rc,Bracaetal-Pageconsensus,MouraLDGauss,MouraLDnonGauss,MouraLDnoisy,TsitsiklisBertsekasAthans,xiao-boyd,BoydGhoshetal,Nedic,Dimakis,KarMoura}) is key to enable continuous learning, e.g., to meet the fundamental requirement of tracking. 
The interplay between adaptation and learning is critical for guaranteeing the successful network operation and to produce reliable inference. 

Such an interplay has been examined rather deeply in the framework of {\em estimation} problems, while less attention has been devoted to {\em detection} problems. Distributed detection over adaptive networks can be succinctly described as follows. A network of dispersed agents, linked together by a given topology, monitors a certain physical phenomenon. As time elapses, the agents gather from the environment streaming data, whose statistical properties depend upon an unknown state of nature, formally represented by a pair of hypotheses, say, ${\cal H}_0$ and ${\cal H}_1$. At {\em each} time instant, {\em each} individual agent must produce a decision inferring the current state of nature, using its own available observations and the local exchange of information obtained from consultation with its neighbors. Due to the requirement of adaptation, the agents must be able to react promptly to drifts in the current state of nature, while guaranteeing an adequate performance level (i.e., low detection error probabilities) at the steady-state, namely, when a given hypothesis is in force for sufficiently long time.

With reference to the above setting, in the recent works~\cite{AdaptiveDetectionArxiv,ADD_ICASSP2014} several interesting results were derived that allowed the authors to establish  fundamental scaling laws for adaptive distributed detection by multi-agent networks. In particular, it was shown there that for diffusion strategies with constant step-size $\mu$, the steady-state error probabilities exhibit an exponential decay, as $\mu\rightarrow 0$, as functions of $1/\mu$. By resorting to a detailed large deviations analysis, the exact scaling law was fully characterized in~\cite{AdaptiveDetectionArxiv,ADD_ICASSP2014} in terms of a rate function $\Phi(\gamma)$ that can be evaluated analytically.
Notably, the results of~\cite{AdaptiveDetectionArxiv,ADD_ICASSP2014} showed that all network agents share the same error exponent so that their error probabilities  are asymptotically equivalent to the leading exponential order as $\mu\rightarrow 0$. This is a remarkable feature of adaptive distributed detection solutions. Readers may consult the introductory remarks of~\cite{AdaptiveDetectionArxiv,ADD_ICASSP2014}, which contain a detailed and motivated summary of these results.

One known limitation of large deviations analysis resides in the fact that it focuses only on the leading exponential order, thus neglecting all sub-exponential terms. The practical implication of this fact can be easily illustrated by a simple example. Assume network agents $1$ and $2$ exhibit asymptotic error probabilities $P_1$ and $P_2$ of the form:
\beq
P_1=e^{-\frac 1 \mu},\qquad P_2=2\,e^{-\frac 1 \mu}=e^{-\frac 1 \mu [1 + o(1)]},
\eeq
where $o(1)$ stands for any correction such that $o(1)\rightarrow 0$ as $\mu\rightarrow 0$. These two probabilities have the same error exponent multiplying $-1/\mu$ (and is equal to one), but the error probability at agent $2$ is always twice that of agent $1$. This is because the factor $2$ is a sub-exponential term. 

The above consideration is particularly relevant in the context of adaptive distributed detection since numerical experiments reported in~\cite{AdaptiveDetectionArxiv,ADD_ICASSP2014} show that, depending on the particular network structure, different agents have different error probabilities, while exhibiting the same scaling law to the leading exponential order. 
Even more remarkably, the experimental results of~\cite{AdaptiveDetectionArxiv,ADD_ICASSP2014}  suggested that the differences in the error probabilities appear to be related to the degrees of connectivity of the various agents. Accordingly, the main purpose of this paper is to perform a refined asymptotic analysis overcoming the limitations of large deviations, in order to capture the fundamental dependence of the error probabilities at different agents upon the specific network connectivity. In this way, we will be able to show that the partial empirical evidences in~\cite{AdaptiveDetectionArxiv,ADD_ICASSP2014} are in fact representative of a {\em universal behavior of distributed detection over adaptive networks}.

Another relevant contribution provided in this work is that, differently from what was done in~\cite{AdaptiveDetectionArxiv,ADD_ICASSP2014}, we shall not limit the analysis to doubly-stochastic combination matrices and will consider instead the broader class of right-stochastic matrices. This latter class is relevant in practical applications. Indeed, while doubly-stochastic matrices might look preferable in detection applications in view of the asymptotic equipartition of the combination weights, they are not always realizable in practice. This happens, for instance, due to physical limitations in the topology and/or communication constraints. 
This generalization in the nature of the combination policy poses new and interesting questions in terms of the interplay between the network topology, the combination weights, and the inference performance. 

In summary, as a result of the refined analysis presented in this work, we will not only be able to evaluate in quantitative terms how network connectivity reflects on the performance of different agents, but we will also be able to clarify how different network structures (i.e., different combination matrices and network topologies) globally impact the network performance.

\subsection{Related Work}
Distributed detection is a classical topic, and the pertinent literature is therefore extensive. Any set of references would be by-no-means exhaustive, so that we refer the reader to~\cite{LongoLookabaughGray,VarshneyBook,ViswanathanVarshney,BlumKassamPoor,ChamberlandVeeravalli,ChamberlandVeeravalliSPMag,ChenTongVarshneySPMag,Saligrama,TsitsiklisAdvSP,WillettSwaszekBlum,LiuSayeed,BarbarossaAcademicPress} as fundamental entry points on the subject. 
In this work we consider fully-decentralized detection problems, i.e., fully-flat network architectures without fusion center, where the agents are only allowed to interact locally. Several recent works address this scenario and, in particular, solutions based on decentralized consensus strategies with decaying step-size have been successfully proposed in~\cite{running-cons,asymptotic-rc,Bracaetal-Pageconsensus,MouraLDGauss,MouraLDnonGauss,MouraLDnoisy}, and the detection performance of these algorithms has been accurately characterized in different asymptotic frameworks~\cite{running-cons,asymptotic-rc,Bracaetal-Pageconsensus,MouraLDGauss,MouraLDnonGauss,MouraLDnoisy}.
However, as already observed in the introduction, a distinguishing feature of our work resides in the emphasis on {\em adaptive} solutions. To enable adaptation, it has been shown that diffusion algorithms with constant step-size are superior to consensus implementations due to an inherent asymmetry in the consensus update that has been shown to be a source of potential instability in the consensus dynamics when constant step-sizes are used~\cite{LopesSayed,CattivelliSayedEstimation,SayedSPmag,SayedProcIEEE}.
Several performance results are already available for diffusion strategies in connection to their mean-square-error (MSE) estimation behavior~\cite{LopesSayed,CattivelliSayedEstimation,SayedSPmag,SayedProcIEEE}. 
The corresponding results for detection applications are relatively limited. In~\cite{CattivelliSayedDetection} the problem of using diffusion algorithms for detection purposes has been considered, with reference to a Gaussian problem. More recently, the general problem of distributed detection over adaptive networks has been posed in~\cite{AdaptiveDetectionArxiv,ADD_ICASSP2014}. By means of a large deviations analysis, the Type-I and Type-II error exponents, for doubly-stochastic combination matrices, were characterized in closed-form in these works. It was shown that the detection performance of the network solution is equivalent to the fully-connected solution to the first-leading order in the exponent. However, numerical and simulation results in~\cite{AdaptiveDetectionArxiv,ADD_ICASSP2014} showed that the network connectivity and the overall structure of the combination weights do matter, and they influence the performance of the individual agents. Analytical formulas for accurate evaluation of the detection error probabilities, and for elucidating the relationship between the network structure and the expected performance, are currently missing. Filling this important gap is the main theme of this work.

\subsection{Summary of Main Result}
In order to introduce the main result, we refer to the steady-state (i.e., as time goes to infinity) output of the diffusion algorithm with constant step-size $\mu$. The steady-state output at the $k$-th agent is denoted by $\by^\star_{k,\mu}$. 
It was shown in~\cite{AdaptiveDetectionArxiv,ADD_ICASSP2014} that, for small step-sizes, $\by^\star_{k,\mu}$ tends to concentrate in the close proximity of the expected value of the local statistics, denoted by $\E[\bx]$. 
In order to characterize the error probabilities, we shall evaluate the probability that $\by^\star_{k,\mu}$ deviates significantly from this expected behavior. Without loss of generality, we shall focus on the following probability:
\beq
\P[\by^\star_{k,\mu} > \gamma],\qquad \gamma>\E[\bx].
\eeq
With reference to the case of doubly-stochastic combination matrices, it was established in~\cite{AdaptiveDetectionArxiv,ADD_ICASSP2014} that
\beq
\lim_{\mu\rightarrow 0} \mu\, \ln  \P[\by^\star_{k,\mu} > \gamma]=
-\Phi(\gamma),
\label{eq:LDPdef0}
\eeq
where $\Phi(\gamma)$ is the so-called {\em rate function}. 
Equivalently, we can rewrite~(\ref{eq:LDPdef0}) as
\beq
\P[\by^\star_{k,\mu}>\gamma]=e^{-\frac{1}{\mu}\,[\Phi(\gamma) + o(1)]}.
\label{eq:expoexpo}
\eeq
This last form of representation highlights the fact that sub-exponential terms are neglected by large deviations analysis. 

A refined analysis can be pursued by seeking an asymptotic approximation, $\mathscr{P}_{k,\mu}(\gamma)$, that ensures the much stronger conclusion:
\beq
\P[\by^\star_{k,\mu}>\gamma]=\mathscr{P}_{k,\mu}(\gamma)[1+o(1)],
\eeq 
i.e.,
\beq
\lim_{\mu\rightarrow 0}\frac{\P[\by^\star_{k,\mu} > \gamma]}{\mathscr{P}_{k,\mu}(\gamma)}=1.
\label{eq:exactasy}
\eeq
This framework is commonly referred to as {\em exact asymptotics}, and has been originally studied in~\cite{BahadurRao} with reference to the simplest case of normalized sums of independent and identically distributed (i.i.d.) random variables --- see also~\cite{Dembo-Zeitouni,DenHollander}.
In the following, we shall often write 
\beq
\P[\by^\star_{k,\mu} > \gamma]\sim\mathscr{P}_{k,\mu}(\gamma)
\eeq
to denote an asymptotic equivalence of the kind~(\ref{eq:exactasy}). It is immediately seen that~(\ref{eq:exactasy}) implies
\beq
\lim_{\mu\rightarrow 0} \mu\, \ln  \P[\by^\star_{k,\mu} > \gamma]=
\lim_{\mu\rightarrow 0} \mu\, \ln  \mathscr{P}_{k,\mu}(\gamma),
\eeq
which means that any exact asymptotic $\mathscr{P}_{k,\mu}(\gamma)$ is able to reflect the leading exponential term.

The main result established in this paper can now be formally stated as follows:
\beq
\boxed{
\mathscr{P}_{k,\mu}(\gamma)
=\sqrt{\frac{\mu}{2\pi \theta_{\gamma}^2\,\phi^{\prime\prime}(\theta_{\gamma})}}
e^{-\frac{1}{\mu}\left[\Phi(\gamma) + \epsilon_{k,\mu}(\theta_{\gamma})\right]}
}
\label{eq:mainPkmu}
\eeq
The main quantities necessary to evaluate the asymptotic approximation $\mathscr{P}_{k,\mu}(\gamma)$ in the above expression are now briefly introduced. 
The function $\phi(t)$ will be described in closed-form, and depends on the underlying statistical model through the moment generating function of the local statistics, and on the network topology only through the {\em limiting} vector of combination weights (aka the Perron eigenvector, see, e.g.,~\cite{SayedProcIEEE}). The quantity $\theta_{\gamma}$ is the solution to the equation:
\beq
\phi^{\prime}(\theta_{\gamma})=\gamma.
\label{eq:stateqsol0}
\eeq
The function $\Phi(\gamma)$, which was the main object of study in~\cite{AdaptiveDetectionArxiv,ADD_ICASSP2014}, is the so-called {\em rate function}, and is computed directly from $\phi(t)$, namely, it is the Fenchel-Legendre transform of $\phi(t)$~\cite{Dembo-Zeitouni,DenHollander}. 

Finally, the (sub-exponential) correction term $\epsilon_{k,\mu}(\theta_{\gamma})$ depends on the underlying statistical model through the moment generating function of the local statistics, and on the network topology through the {\em actual} (not only the limiting) network combination weights. This correction satisfies:
\beq
\epsilon_{k,\mu}(\theta_{\gamma})\stackrel{\mu\rightarrow 0}{\longrightarrow} 0.
\label{eq:epsiloncorr}
\eeq
In particular, the ratio $\epsilon_{k,\mu}(\theta_{\gamma})/\mu$ stays bounded as $\mu\rightarrow 0$. Loosely speaking, this means that the overall correction appearing at the exponent in~(\ref{eq:mainPkmu}) plays, asymptotically, the role of a {\em constant} correction.

Despite its apparent complexity, Eq.~(\ref{eq:mainPkmu}) possesses a well defined structure, revealing important connections with the physical behavior of the adaptive distributed system under consideration. Let us elucidate some of these features.
We start by the leading order in the exponent. It is easy to see that:
\beqa
\mu \ln {\mathscr{P}_{k,\mu}(\gamma)}&=&-\Phi(\gamma)-\epsilon_{k,\mu}(\theta_{\gamma})+\nonumber\\
&&
\frac \mu 2 \ln\mu -
\frac \mu 2\ln\left[2\pi \theta_{\gamma}^2\,\phi^{\prime\prime}(\theta_{\gamma})\right] 
\nonumber\\
&\stackrel{\mu\rightarrow 0}{\longrightarrow}& -\Phi(\gamma),
\eeqa
which easily follows from~(\ref{eq:epsiloncorr}).
This means that the approximation $\mathscr{P}_{k,\mu}$ in~(\ref{eq:mainPkmu}) can be regarded as
\beq
\mathscr{P}_{k,\mu}(\gamma)=e^{-\frac{1}{\mu}[\Phi(\gamma) + o(1)]}.
\label{eq:Pkmualternativeform}
\eeq
This is consistent with result~(\ref{eq:LDPdef0}), which was established in~\cite{AdaptiveDetectionArxiv,ADD_ICASSP2014} for the case of doubly-stochastic combination matrices.
The terms of order $o(1)$ in~(\ref{eq:Pkmualternativeform}) collect all the sub-exponential corrections that appear in~(\ref{eq:mainPkmu}). They can be separated into two categories. 
The first correction is the term $\sqrt{\frac{\mu}{2\pi \theta_{\gamma}^2\,\phi^{\prime\prime}(\theta_{\gamma})}}$ in~(\ref{eq:mainPkmu}). This term is a typical sub-exponential refinement arising in the framework of exact asymptotics, and is a consequence of a local Central Limit Theorem (see~\cite{BahadurRao,Dembo-Zeitouni,DenHollander}).
Observe that this correction, which is related to the network topology only through the Perron eigenvector, is {\em independent} of the agent index $k$, and is therefore applicable to all agents. In contrast, the second correction $\epsilon_{k,\mu}(\theta_{\gamma})$ depends on the agent index $k$ and, as it will be detailed in the exact statement of the main theorems, takes into account the entire network topology and combination weights.  

\vspace*{5pt}
The above considerations lead to the important conclusion that~(\ref{eq:mainPkmu}) provides a detailed and revealing assessment of the {\em universal behavior of distributed detection over adaptive networks}: as functions of $1/\mu$, the error (log-)probability curves corresponding to different agents not only stay nearly-parallel to each other, but they are also ordered following a criterion dictated by the correction term $\epsilon_{k,\mu}(\theta_\gamma)$. As we shall see later --- see Fig.~\ref{fig:fig1} for an example, this criterion reflects the degree of connectivity of each agent. Depending on the combination weights, the more connected an agent is, the lower its error probability will be, and the correction term $\epsilon_{k,\mu}(\theta_\gamma)$ is sufficiently rich to capture this behavior.

\vspace*{5pt}
\noindent
{\bf Notation.} We use boldface letters to denote random variables, and normal font letters for their realizations. Capital letters refer to matrices, small letters to both vectors and scalars. 
Sometimes we violate this latter convention, for instance, we denote the total number of sensors by $S$. The symbols $\P$ and $\E$ are used to denote the probability and expectation operators, respectively. 
The notation $\P_h$ and $\E_h$, with $h=0,1$, means that the pertinent statistical distribution corresponds to hypothesis ${\cal H}_0$ or ${\cal H}_1$.

\section{Problem formulation}

We consider a sensor network that collects observations about a physical phenomenon of interest. Data are assumed to be spatially and temporally independent and identically distributed.
From the observation measured at time $n$, the $k$-th agent computes its local statistic (the observation itself, or a suitable function thereof), which is denoted by $\bx_k(n)$, $k=1,2,\dots,S$. The mean and variance of $\bx_k(n)$ will be denoted by $\E[\bx]$ and  $\sigma_x^2$, respectively.  
To avoid trivialities, throughout the paper it is assumed that the random variable $\bx_k(n)$ is non-degenerate.

\subsection{Diffusion Strategy}
Following the framework developed in~\cite{AdaptiveDetectionArxiv,ADD_ICASSP2014}, we focus on the class of diffusion strategies for adaptation over networks~\cite{CattivelliSayedDetection,SayedSPmag,SayedProcIEEE}, and in particular on the ATC (Adapt-Then-Combine) implementation due to some inherent advantages in terms of a slightly improved mean-square-error performance relative to other forms~\cite{SayedSPmag}. Extensions to other diffusion implementations, as well as to consensus implementations, are certainly possible. In the ATC algorithm, each node $k$ updates its state from $\by_{k}(n-1)$ to $\by_{k}(n)$ through local cooperation with its neighbors as follows:

\begin{eqnarray}
\bv_k(n)&=&\by_k(n-1) + \mu[\bx_k(n)-\by_k(n-1)],\label{eq:diff1}\\
\by_k(n)&=&\sum_{\ell=1}^S a_{k,\ell} \bv_{\ell}(n),
\label{eq:diff2}
\end{eqnarray}
where $0<\mu\ll 1$ is a small step-size parameter. It is seen that node $k$ first uses its locally available statistic $\bx_k(n)$, to update its state from $\by_{k}(n-1)$ to an intermediate value $\bv_k(n)$. The other network agents simultaneously perform similar updates using their local statistics. Subsequently, node $k$ aggregates the intermediate states of its neighbors using nonnegative convex combination weights $\{a_{k,\ell}\}$ that add up to one. Again, all other network agents perform a similar calculation. Collecting the combination weights into a square matrix~$A=[a_{k,\ell}]$, then $A$ is a right-stochastic matrix, namely, the entries on each row add up to one. Formally:
\beq
a_{k,\ell}\geq 0,\quad A\mathds{1}=\mathds{1},
\eeq
with $\mathds{1}$ being a column-vector with all entries equal to $1$.
We denote the $n$-th power of $A$ by
\beq
B_n=[b_{k,\ell}(n)]\dfz A^n.
\eeq
Throughout this article, we assume that $A$ has second largest eigenvalue magnitude strictly less than one, which yields~\cite{xiao-boyd,Johnson-Horn}:
\beq
\boxed{
b_{k,\ell}(n)\stackrel{n\rightarrow\infty}{\longrightarrow} p_{\ell}
\quad 
\Leftrightarrow
\quad 
B_n\stackrel{n\rightarrow\infty}{\longrightarrow} \mathds{1} p
}
\label{eq:bconv}
\eeq
where the limiting combination weights $\{p_\ell\}$ satisfy, for all $\ell=1,2,\dots,S$:
\beq
\boxed
{
p A=p,\quad p_\ell>0,\quad \sum_{\ell=1}^S p_\ell=1
}
\label{eq:stateq}
\eeq
and the row vector $p=[p_1,p_2,\dots,p_S]$ is usually referred to as the Perron eigenvector of $A$ --- see, e.g.,~\cite{SayedProcIEEE}.
We remark that the condition on $A$ is automatically satisfied by network topologies that are strongly-connected~\cite{SayedProcIEEE}, i.e., when there is always a path with nonzero weights between any pair of nodes, and at least one node in the network has a self-loop ($a_{k,k}>0$ for some agent $k$).

\subsection{Steady-State Distribution}
In order to design and characterize an inference system based upon the sensor output $\by_k(n)$, knowledge of the distribution of $\by_k(n)$ is crucial. However this knowledge is seldom available, except for very special cases (e.g., Gaussian observations). 
A common and well-established approach in the adaptation literature~\cite{Sayed2008adaptive,SayedSPmag} to address this difficulty is to focus on $i)$ the steady-state properties (as $n\rightarrow\infty$), and $ii)$ the small step-size regime ($\mu\rightarrow 0$). Accordingly, throughout the paper, the term {\em steady-state} refers to the limit as the time-index $n$ goes to infinity, while the term {\em asymptotic} refers to the slow adaptation regime  where $\mu\rightarrow 0$. 

We start by considering the steady-state behavior of $\by_k(n)$ for a given step-size $\mu$. To this aim, it is useful to recast the pair of equations given by~(\ref{eq:diff1}) and~(\ref{eq:diff2}) into the following single equation, which is obtained by straightforward algebra:  
\beqa
\by_k(n)&=&
\underbrace{(1-\mu)^n\,\sum_{\ell=1}^S b_{k,\ell} (n) \by_\ell(0)}_{\textnormal{transient term}}+
\nonumber\\
&& 
\sum_{i=1}^{n} \sum_{\ell=1}^S \mu (1-\mu)^{i-1} b_{k,\ell}(i) \bx_\ell(n-i+1).
\nonumber\\
\label{eq:mainATCrecast}
\eeqa
Since the random variables $\bx_k(n)$ are i.i.d. across time, and since we shall be only concerned with the distribution of partial sums involving these terms, it is convenient to define the following random variable: 
\beq
\by^\star_k(n)\dfz
\sum_{i=1}^n \sum_{\ell=1}^S  \mu (1-\mu)^{i-1}  b_{k,\ell}(i) \bx_{\ell}(i),
\label{eq:finitehorizykn}
\eeq
which {\em shares the same distribution of} the second term on the RHS of~(\ref{eq:mainATCrecast}) --- see also the discussion in~\cite{AdaptiveDetectionArxiv,ADD_ICASSP2014}. Formally:
\beq
\by^\star_k(n)
\stackrel{d}{=}
\sum_{i=1}^{n} \sum_{\ell=1}^S \mu (1-\mu)^{i-1} b_{k,\ell}(i) \bx_\ell(n-i+1),
\label{eq:ystarknnulltransient}
\eeq
where the notation $\stackrel{d}{=}$ denotes equality {\em in distribution}.

In Theorem~$1$ of~\cite{AdaptiveDetectionArxiv,ADD_ICASSP2014}, the existence of a steady-state random variable characterizing the diffusion output has been established. This can be summarized by the following statement (the symbol $\rightsquigarrow$ means convergence in distribution):
\beq
\boxed{
\by_k(n)\stackrel{n\rightarrow\infty}{\rightsquigarrow} \by^\star_{k,\mu}\dfz
\sum_{i=1}^{\infty} \sum_{\ell=1}^S 
\mu\,(1-\mu)^{i-1} b_{k,\ell}(i) \bx_\ell(i)
}
\label{eq:Theo1}
\eeq
where the first two moments of $\by^\star_{k,\mu}$ are given by
\beqa
\E[\by^\star_{k,\mu}]&=&\E[\bx],
\label{eq:expectatystar}
\\
\VAR[\by^\star_{k,\mu}]&=&
\sigma^2_x \, \sum_{i=1}^\infty \sum_{\ell=1}^S  \mu^2(1-\mu)^{2(i-1)}  b^2_{k,\ell}(i).
\label{eq:varystar}
\eeqa
Actually, the proof in~\cite{AdaptiveDetectionArxiv,ADD_ICASSP2014} focused on the case of doubly-stochastic combination matrices, but it is immediate to verify that the same argument holds for right-stochastic combination matrices and leads to~(\ref{eq:Theo1})--(\ref{eq:varystar}).

\subsection{The Inference Problem}
In the distributed detection formulation, each agent in the network must perform a binary hypothesis test, in an adaptive and fully decentralized manner. 
In this setting, the agents in the network continually collect an increasing amount of streaming data, whose statistical properties depend upon an {\em unknown} binary state of nature, which is represented by a pair of hypotheses, say, ${\cal H}_0$ and ${\cal H}_1$.
The statistics $\bx_k(n)$ are spatially and temporally i.i.d., {\em conditioned} on the hypothesis that gives rise to them.
In what follows, we shall always assume that $\E_0[\bx]\neq\E_1[\bx]$. To get some intuition about the meaning of this condition, consider that, when the local statistic $\bx$ is chosen as the log-likelihood of the local observation, the inequality $\E_0[\bx]\neq\E_1[\bx]$ is simply a way to state that the detection problem is {\em identifiable}, namely, that it is not singular\cite{poorbook}. Furthermore, and without loss of generality, we assume that:
\beq
\E_0[\bx]<\E_1[\bx].
\eeq
At time $n$, the $k$-th sensor needs to produce a decision about the state of nature, based upon its state value $\by_k(n)$.
As discussed in~\cite{AdaptiveDetectionArxiv,ADD_ICASSP2014}, the computation of $\by_k(n)$ via algorithm~(\ref{eq:diff1}) and~(\ref{eq:diff2}) is motivated by the fact that this implementation essentially results in a value $\by_k(n)$ that corresponds to a weighted average of the local statistics $\bx_k(n)$. Such additive constructions for the state variable are not only convenient from an implementation point of view, but they are also meaningful from a theoretical standpoint. Indeed, in the classical, centralized and non-adaptive theory with i.i.d. data (where the optimal detection statistic is the sum of the log-likelihoods), as well as in more general  frameworks such as locally optimum detection or universal hypothesis testing~\cite{kassam,poorbook}, using additive detection statistics is the best choice. 
Moreover, the decision regions employed by the detector are often in the form of single-threshold rules, which, for a variety of detection problems, exhibit  asymptotic optimality properties also in our adaptive and distributed setting, as already shown in~\cite{AdaptiveDetectionArxiv,ADD_ICASSP2014}. 
Motivated by these considerations, in this work the test implemented by agent $k$ at time $n$ is therefore chosen to be of the form:
\beq
\by_k(n)\mathop{\lesseqgtr}_{{\cal H}_1}^{{\cal H}_0} \gamma.
\label{eq:TheTest}
\eeq 
The performance of the test is typically expressed in terms of the Type-I (choose ${\cal H}_1$ when ${\cal H}_0$ is true) and Type-II (choose ${\cal H}_0$ when ${\cal H}_1$ is true) error probabilities.
With reference to the steady-state performance, the Type-I and Type-II error probabilities are defined as, respectively:
\beq
\alpha_{k,\mu}\dfz\P_0[\by^\star_{k,\mu} >\gamma],
\qquad
\beta_{k,\mu}\dfz\P_1[\by^\star_{k,\mu}\leq \gamma].
\label{eq:alphabet3}
\eeq

\section{Main Theorems}
\label{sec:maintheorems}
We now introduce the basic quantities necessary to characterize the system performance. As we shall see, it is important to understand the behavior of the Logarithmic Moment Generating Function (LMGF) of the steady-state variable $\by^\star_{k,\mu}$. In Appendix A we state and prove a theorem (Theorem 1) that establishes several useful properties of the LMGF, of its derivatives, and of its limiting behavior as $\mu\rightarrow 0$. In the following description we shall mention and use some of these properties, referring the reader to Appendix A for a more detailed explanation.
\begin{itemize}
\item
{\em Local LMGF}. 
\\
\noindent
The LMGF of the local statistics $\bx_k(n)$ is defined as:
\beq
\psi(t)\dfz \ln \E[e^{t \bx_k(n)}],
\eeq
and is independent of $k$. A fundamental role in our results will be played by the following averaged version of the LMGF $\psi(t)$, which has already been used in~\cite{AdaptiveDetectionArxiv,ADD_ICASSP2014}:
\beq
\omega(t)\dfz\int_{0}^{t}\frac{\psi(\tau)}{\tau}d\tau.
\label{eq:omegadef}
\eeq
\item
{\em Steady-state LMGF}. 
\\
\noindent
The LMGF of the steady-state variable $\by^\star_{k,\mu}$ is defined as:
\beq
\phi_{k,\mu}(t)\dfz \ln \E[e^{t \by_{k,\mu}^\star}],
\eeq
and admits the following representation [Theorem 1, Eq.~(\ref{eq:Theorem1_1})]:
\beq
\phi_{k,\mu}(t)=\sum_{i=1}^\infty\sum_{\ell=1}^S \psi\left(\mu (1-\mu)^{i-1}b_{k,\ell}(i) t\right).
\label{eq:phikmurepresent}
\eeq
\item
{\em Limiting properties of the steady-state LMGF}. 
\\
\noindent
We will be primarily concerned with the limiting normalized LMGF:
\beq
\phi(t)\dfz
\lim_{\mu\rightarrow 0} \mu\,
\phi_{k,\mu}(t/\mu),
\label{eq:firstdefoflimLMGF}
\eeq
where the above limit exists, and is given by [Theorem 1, Eq.~(\ref{eq:Theorem1_2})]:
\beq
\boxed{
\phi(t)=\sum_{\ell=1}^S 
\omega(p_\ell t)=\sum_{\ell=1}^S \int_{0}^{p_\ell t}\frac{\psi(\tau)}{\tau}d\tau
}
\label{eq:phitfirstclaim}
\eeq
Equation~(\ref{eq:phitfirstclaim}) emphasizes that $\phi(t)$ depends on the underlying statistical model through the LMGF $\psi(t)$ of the local statistics $\bx_k(n)$, and on the network topology only through the Perron eigenvector $p$, i.e., on the limiting combination weights. 
For doubly-stochastic matrices we have $p_{\ell}=1/S$, so that the above formula gives
\beq
\phi(t)=S\omega(t/S)=S\int_{0}^{t/S}\frac{\psi(\tau)}{\tau}d\tau,
\label{eq:phifordoublystoch}
\eeq
which is consistent with the results of~\cite{AdaptiveDetectionArxiv,ADD_ICASSP2014}. 


It is of interest to consider an alternative representation for~(\ref{eq:phitfirstclaim}). To this aim, we introduce the LMGF of the averaged random variable $\sum_{\ell=1}^S p_\ell \,\bx_\ell(n)$, namely:
\beq
\bar \psi(t)\dfz\sum_{\ell=1}^S \psi(p_\ell t),
\eeq
and by straightforward calculations we get from~(\ref{eq:phitfirstclaim}): 
\beq
\boxed{
\phi(t)=\int_{0}^t \frac{\bar\psi(\tau)}{\tau}d\tau
}
\eeq
\item
{\em Property of the LMGF derivatives}. 
\\
\noindent
In Theorem 1, part $ii)$, we establish that the derivatives of $\phi_{k,\mu}(t)$ and of the limiting function $\phi(t)$ can be evaluated by interchanging the differential and limit operators, yielding, in particular [Theorem 1, Eq.~(\ref{eq:Theorem1_3})]:
\beqa
\phi_{k,\mu}^{\prime}(t)
&=&
\mu \sum_{i=1}^\infty\sum_{\ell=1}^S (1-\mu)^{i-1}b_{k,\ell}(i)  \times\nonumber\\
&&
\psi^\prime\left(\mu (1-\mu)^{i-1}b_{k,\ell}(i) t\right),
\label{eq:phikmuderrepresent}
\eeqa
and [Theorem 1, Eq.~(\ref{eq:Theorem1_4})]
\beq
\lim_{\mu\rightarrow 0} \phi_{k,\mu}^{\prime}(t/\mu)=\phi^\prime(t)=\frac 1 t\sum_{\ell=1}^S \psi(p_\ell t).
\label{eq:firstderLMGFconv}
\eeq
\item
{\em Convergence errors}. 
\\
\noindent
As we shall see, the convergence errors corresponding to~(\ref{eq:firstdefoflimLMGF}) and~(\ref{eq:firstderLMGFconv}) will play an important role in characterizing the error probabilities. The following rates of convergence  [Theorem 1, Eq.~(\ref{eq:Theorem1_6})] hold:
\beqa
\phi(t)-\mu\,\phi_{k,\mu}(t/\mu)&=&{\cal O}(\mu)
\label{eq:firstconverr}
\\
\phi^\prime(t)-\phi^\prime_{k,\mu}(t/\mu)&=&{\cal O}(\mu),
\label{eq:secondconverr}
\eeqa
where the notation $f_\mu={\cal O}(\mu)$ means that the ratio $f_\mu/\mu$ stays bounded as $\mu\rightarrow 0$.

\item
{\em Fenchel-Legendre transforms}. 
\\
\noindent
The Fenchel-Legendre transform of the function $\phi(t)$ is defined by~\cite{Dembo-Zeitouni,DenHollander}:
\beq
\Phi(\gamma)\dfz\sup_{t\in\mathbb{R}}[\gamma t -\phi(t)].
\label{eq:FLtransf}
\eeq
We shall use capital letters to denote Fenchel-Legendre transforms, as done in~(\ref{eq:FLtransf}). 
It is further useful to introduce the domain where $\Phi(\gamma)$ is finite, namely:
\beq
{\cal D}_{\Phi}=\{\gamma\in\mathbb{R}:\;\Phi(\gamma)<\infty\}.
\eeq
The notation ${\cal D}^o_{\Phi}$ adopted in the sequel will denote the interior of the set ${\cal D}_{\Phi}$. 
\end{itemize}

\vspace*{10pt}
\noindent
We can now state our second theorem, which generalizes Theorem 3 from~\cite{AdaptiveDetectionArxiv,ADD_ICASSP2014} to handle the case of right-stochastic matrices. 

\vspace*{5pt}
\noindent
{\bf \textsc{Theorem 2} (Large deviations of $\by^\star_{k,\mu}$ as $\bm{\mu\rightarrow 0}$, for right-stochastic matrices).}
{\em
Assume that $\psi(t)<+\infty$ for all $t\in \mathbb{R}$. 
Then, for all $k=1,2,\dots,S$:
\begin{itemize}
\item[$i)$]
The steady-state variable $\by^\star_{k,\mu}$ obeys a Large Deviation Principle (LDP) with rate function given by the Fenchel-Legendre transform $\Phi(\gamma)$ defined by~(\ref{eq:FLtransf}), namely, for any Borel measurable region $\Gamma\in\mathbb{R}$:
\beq
\lim_{\mu\rightarrow 0} \mu\, \ln  \P[\by^\star_{k,\mu}\in \Gamma]=
-\inf_{\gamma\in \Gamma} \Phi(\gamma).
\label{eq:LDPdef}
\eeq
\item[$ii)$]
$\phi^{\prime\prime}(t)>0$ for all $t\in\mathbb{R}$, implying that $\phi(t)$ is strictly convex. Moreover, the rate function $\Phi(\gamma)$ is strictly convex in ${\cal D}^{o}_{\Phi}$, and attains its unique minimum at $\gamma=\E[\bx]$, with $\Phi(\E[\bx])=0$. 
\end{itemize}
}

\vspace*{5pt}
\noindent
{\em Proof:}  By Theorem 1, part $i)$, Appendix A, we can conclude that the function $\phi(t)$ in~(\ref{eq:firstdefoflimLMGF}) exists and is finite for all $t\in\mathbb{R}$. Since, by definition, $\phi(t)$ is a normalized limiting LMGF, claim $i)$ of the present theorem follows by application of the G\"artner-Ellis Theorem~\cite{Dembo-Zeitouni,DenHollander}. 

Claim $ii)$ is a direct extension to the case of right-stochastic combination matrices of the results proved for doubly-stochastic combination matrices in Appendix C of~\cite{AdaptiveDetectionArxiv}.

~\hfill$\square$

\vspace*{10pt}
\noindent
According to property $i)$ in Theorem 2, the steady-state random variable $\by^\star_{k,\mu}$ obeys a LDP with rate function $\Phi(\gamma)$. In view of the convexity properties of $\Phi(\gamma)$ stated in $ii)$, for any $\gamma\in{\cal D}^{o}_{\Phi}$, with $\gamma>\E[\bx]$, the infimum over an interval  of the form $(\gamma,\infty)$ always lies on the boundary point $\gamma$. The same conclusion is reached for the infimum over an interval of the form $(-\infty,\gamma)$ when $\gamma<\E[\bx]$. Formally, for any $\gamma\in{\cal D}^{o}_{\Phi}$, we have:
\beqa
\lim_{\mu\rightarrow 0} \mu\, \ln  \P[\by^\star_{k,\mu}>\gamma]&=&
-\Phi(\gamma),\qquad \gamma>\E[\bx]
\label{eq:LDP1}\\
\lim_{\mu\rightarrow 0} \mu\, \ln  \P[\by^\star_{k,\mu}\leq\gamma]&=&
-\Phi(\gamma),\qquad \gamma<\E[\bx].
\label{eq:LDP2}
\eeqa
Note also that, according to Theorem 2, part $ii)$, the choice $\gamma=\E[\bx]$ yields $\Phi(\gamma)=\Phi(\E[\bx])=0$, that is, a null exponent. This is the uninteresting case where the error probability does {\em not} vanish exponentially. Accordingly, this situation will be ruled out from the forthcoming analysis. 
Relationships~(\ref{eq:LDP1}) and~(\ref{eq:LDP2}), along with the latter observation, have an immediate implication as regards the choice of the detection threshold. Indeed, in the light of~(\ref{eq:alphabet3}), one must ensure that 
\beq
\E_0[\bx]<\gamma<\E_1[\bx],
\eeq 
in order to guarantee the exponential decay of both error probabilities $\alpha_{k,\mu}$ and $\beta_{k,\mu}$.

\vspace*{10pt}
The exact asymptotics of $\by^\star_{k,\mu}$ are now characterized in the following theorem.

\vspace*{5pt}
\noindent
{\bf \textsc{Theorem 3} (Exact asymptotics of  $\by^\star_{k,\mu}$ as $\bm{\mu\rightarrow 0}$).}
{\em 
Assume that $\bx_k(n)$ is not lattice, and that $\psi(t)<+\infty$ for all $t\in \mathbb{R}$. 
Let $\gamma\in{\cal D}^o_{\Phi}$, with $\gamma>\E[\bx]$, and let $\theta_{\gamma}$ be the unique solution to the stationary equation:
\beq
\phi^{\prime}(\theta_{\gamma})=\gamma.
\label{eq:stateqsol}
\eeq
Then, $\theta_{\gamma}>0$ and, for $k=1,2,\dots,S$:
\beq
\P[\by^\star_{k,\mu} > \gamma]\sim\mathscr{P}_{k,\mu}(\gamma),
\label{eq:exactasy_theorem}
\eeq
where
\beq
\boxed{
\mathscr{P}_{k,\mu}(\gamma)
\dfz
\sqrt{\frac{\mu}{2\pi \theta_{\gamma}^2\,\phi^{\prime\prime}(\theta_{\gamma})}}
e^{-\frac{1}{\mu}\left[\Phi(\gamma) + \epsilon_{k,\mu}(\theta_{\gamma})\right]}
}
\label{eq:mainPkmu_theorem}
\eeq
with
\begin{subequations}
\beq
\epsilon_{k,\mu}(t)=
\phi(t)-\mu \phi_{k,\mu}(t/\mu).
\label{eq:epsilonkmumainexpression}
\eeq
The correction term $\epsilon_{k,\mu}(t)$ can be refined to:
\beq
\epsilon_{k,\mu}(t)=
[\phi(t)-\mu \phi_{k,\mu}(t/\mu)]
+\frac{[\phi^\prime(t)-\phi^\prime_{k,\mu}(t/\mu)]^2}{2\,\phi^{\prime\prime}(t)}.
\label{eq:epsilonkmumainexpression_bis}
\eeq
\end{subequations}
}

\vspace*{5pt}
\noindent
{\em Proof:} See Appendix B.

~\hfill$\square$

\noindent
\textsc{Remark I}. The key ingredients to computing $\mathscr{P}_{k,\mu}(\gamma)$ in~(\ref{eq:mainPkmu_theorem}) are the moment generating function $\psi(t)$ of the local statistics $\bx_k(n)$, and the combination matrix $A$.
Indeed, the quantities $\Phi(\gamma)$ and $\theta_{\gamma}$ depend on the function $\phi(t)$, which in turn depends on $\psi(t)$ and on the {\em limiting} combination weights $p_\ell$. The correction term $\epsilon_{k,\mu}(t)$ (in both versions) depends on $\psi(t)$ and on the {\em actual} combination weights $b_{k,\ell}(i)$. 

~\hfill$\square$

\noindent
\textsc{Remark II}. It is useful to comment on the reason for reporting two expressions for $\epsilon_{k,\mu}(t)$. 
First, note that, from~(\ref{eq:secondconverr}), the quantity
\beq
\frac{1}{\mu} \frac{[\phi^\prime(t)-\phi^\prime_{k,\mu}(t/\mu)]^2}{2\,\phi^{\prime\prime}(t)}=
\frac{\mu}{{2\,\phi^{\prime\prime}(t)}} \left[\frac{\phi^\prime(t)-\phi^\prime_{k,\mu}(t/\mu)}{\mu}\right]^2
\eeq
vanishes as $\mu\rightarrow 0$, implying that, asymptotically, using~(\ref{eq:epsilonkmumainexpression}) or~(\ref{eq:epsilonkmumainexpression_bis}) in~(\ref{eq:mainPkmu_theorem}) makes no difference.
This equivalence is not surprising because, in principle, one can construct an arbitrary number of {\em distinct} approximations that are {\em asymptotically} equivalent. 
The relevant fact is that different forms for the correction correspond to different approximations of the true error probability, which will perform differently for {\em finite} values of $1/\mu$. 
We shall see in the proof of Theorem 3 why~(\ref{eq:epsilonkmumainexpression_bis}) is a refined version of~(\ref{eq:epsilonkmumainexpression}). From now on, unless otherwise stated, we shall make reference to~(\ref{eq:epsilonkmumainexpression_bis}).

~\hfill$\square$

\noindent
\textsc{Remark III}.  Applying~(\ref{eq:firstconverr}) and~(\ref{eq:secondconverr}), it is immediately seen that:
\beq
\epsilon_{k,\mu}(\theta_{\gamma})={\cal O}(\mu),
\eeq
meaning that the overall correction appearing at the exponent in~(\ref{eq:mainPkmu_theorem}) remains bounded as $\mu\rightarrow 0$. 

~\hfill$\square$

\noindent
\textsc{Remark IV}. The theorem is stated with reference to large deviations of the type $\P[\by^\star_{k,\mu}>\gamma]$, with $\gamma>\E[\bx]$. We notice that the main formulas~(\ref{eq:mainPkmu_theorem}), (\ref{eq:epsilonkmumainexpression}) and~(\ref{eq:epsilonkmumainexpression_bis}), for the complementary case $\gamma<\E[\bx]$ and $\P[\by^\star_{k,\mu}\leq\gamma]$ remain unchanged. The only difference in the claim is that $\theta_\gamma<0$. These facts can be verified by repeating the proof of the theorem for this complementary case. Alternatively, they can be verified by directly applying the claim of Theorem 3 to the random variable $-\by^\star_{k,\mu}$, and by observing that the difference between $<$ and $\leq$ is immaterial in the proof of the theorem. 

~\hfill$\square$

\noindent
\textsc{Remark V}.
The proof of Theorem 3 assumes that the local statistic $\bx_k(n)$ is not of lattice type. We recall that the distribution of a lattice random variable with span $\Delta$ is concentrated at the points $d,d\pm\Delta,d\pm 2\Delta,\dots$ for a certain real number $d$ --- see, e.g.,~\cite{FellerBookV2}. The use of the non-lattice assumption in our proof is discussed in the Remark VI at the end of Appendix B. It is useful to observe that, when dealing with the classical and simplest case of normalized sums of i.i.d. random variables, the exact asymptotics for the lattice case must take into account a further correction term that is related to the lattice span~\cite{Dembo-Zeitouni}. However, as already observed, our asymptotic setting with vanishing step-size is markedly different from that addressed in the literature for the problem of normalized sums of i.i.d. random variables. It is therefore not easy to anticipate if the results of Theorem 3 hold as they are also for the lattice case, if they hold as they are under additional conditions, or if further corrections are needed.

~\hfill$\square$

Before ending this section it is useful to collect the steps needed to evaluate~(\ref{eq:mainPkmu_theorem}).
Depending on the particular application, the formulas in this listing might need to be evaluated numerically. For practical purposes, the infinite summations appearing in $\epsilon_{k,\mu}(\cdot)$ must be obviously truncated.   

\vspace*{30pt}
\hrule
\hrule
\vspace*{5pt}
{\bf 
Evaluation of $\mathscr{P}_{k,\mu}(\gamma)$ in~(\ref{eq:mainPkmu_theorem}).
}
\vspace*{5pt}
\hrule
\hrule
\vspace*{5pt}
\begin{itemize}
\item[{\bf 1)}]
Find the solution $\theta_{\gamma}$ to the stationary equation:
\[
\phi^{\prime}(\theta_{\gamma})=\frac{1}{\theta_\gamma}\sum_{\ell=1}^S
\psi(p_\ell \theta_{\gamma})=\gamma.
\]
\item[{\bf 2)}]
Compute the rate function:
\[
\Phi(\gamma)=\sup_{t\in\mathbb{R}}[\gamma t-\phi(t)]=\gamma\theta_{\gamma} - \phi(\theta_{\gamma}).
\]
\item[{\bf 3)}]
Compute the second derivative:
\[
\phi^{\prime\prime}(\theta_{\gamma})=\frac{1}{\theta^2_{\gamma}}\sum_{\ell=1}^S [(p_\ell \theta_{\gamma})\psi^{\prime}(p_\ell \theta_{\gamma})-\psi(p_\ell \theta_{\gamma})].
\]
\item[{\bf 4)}]
Compute the convergence errors~(\ref{eq:firstconverr}) and~(\ref{eq:secondconverr}) by using~(\ref{eq:phikmurepresent}) and~(\ref{eq:phikmuderrepresent}):
\beqa
{\cal C}_1&=&
\phi(\theta_{\gamma})-\mu\,\sum_{i=1}^\infty\sum_{\ell=1}^S \psi\left((1-\mu)^{i-1}b_{k,\ell}(i)\theta_{\gamma}\right),\nonumber
\\
{\cal C}_2&=&\phi^{\prime}(\theta_\gamma)-
\mu \sum_{i=1}^\infty\sum_{\ell=1}^S (1-\mu)^{i-1}b_{k,\ell}(i)  \times\nonumber\\
&&
\psi^\prime\left((1-\mu)^{i-1}b_{k,\ell}(i) \theta_{\gamma}\right).\nonumber
\eeqa
\item[{\bf 5)}]
Compute the correction term in~(\ref{eq:epsilonkmumainexpression_bis}):
\[
\epsilon_{k,\mu}(\theta_{\gamma})=
{\cal C}_1+\frac{{\cal C}_2^2}{2\,\phi^{\prime\prime}(\theta_\gamma)}.
\]
\item[{\bf 6)}]
Using the above quantities, evaluate $\mathscr{P}_{k,\mu}(\gamma)$.
\end{itemize}
\vspace*{5pt}
\hrule
\hrule
\vspace*{5pt}

%
%
%

\section{Importance Sampling and Cram\'er's transform}
\label{sec:ISMC}

Real-world detectors are usually required to be high-performing, i.e., they must exhibit very low error probabilities. Unfortunately,  standard Monte Carlo techniques to estimate their performance become unfeasible for error probability values in the order of $10^{-6}$. 
To overcome this issue, we shall resort to {\em importance sampling} techniques~\cite{BucklewBook}, which can dramatically reduce the number of runs needed to reach a prescribed level of estimation accuracy.  We start by illustrating briefly the importance sampling philosophy. Then, in the next section, we show how it should be applied to the adaptive distributed detection problem. 

\subsection{Cram\'er's Transform}
Let us refer to a random variable $\by$, assumed continuous for ease of description. We shall denote by $f(y)$ its probability density function (pdf). Consider now another pdf $\tilde f(y)$ that does not vanish (except for zero-measure sets) when $f(y)>0$, and introduce the following weighting function:
\beq
w(y)=\frac{f(y)}{\tilde f(y)},
\label{eq:weightwatchers}
\eeq 
that is, the likelihood ratio between $f(y)$ and $\tilde f(y)$. It then holds that:
\beqa
\P[\by>\gamma]&=&
\int_{\gamma}^{\infty}f(y) dy
=
\int_{\gamma}^{\infty}w(y) \tilde f(y) dy\nonumber\\
&=&
\E_{\tilde f}\left[w(\by) {\cal I}_{\{\by>\gamma\}}\right],
\label{eq:basicOfIs}
\eeqa
where ${\cal I}_{{\cal E}}$ is the indicator of an event ${\cal E}$, and $\E_{\tilde f}[\cdot]$ denotes expectation computed over the {\em transformed} pdf $\tilde f(y)$.
The above equation shows that the quantity to be estimated can be regarded as the expectation, under the transformed pdf, of the  indicator of the event $\{\by>\gamma\}$, {\em weighted} by the function $w(y)$. 
The rationale behind importance sampling is that, by an appropriate choice of the weighting function, it is possible to map an event that is rare under the original sampling pdf $f(y)$, into an event that is {\em not} rare under the new sampling pdf $\tilde f(y)$. In this way, the number of Monte Carlo iterations needed to estimate the expectation is reduced, because (important) samples are generated around the body (not the tail) of the new distribution. An accurate estimate of the probability tails is enabled by the weighting function $w(y)$.  

\vspace*{5pt}
When working with random variables obeying a LDP, there is a classical way to select the transformed pdf $\tilde f(y)$. This is usually referred to as {\em exponential twisting} of $f(y)$, and amounts to selecting~\cite{Dembo-Zeitouni,DenHollander}:
\beq
\tilde f(y)=e^{\eta y-\ln\E[e^{\eta\by}]}\,f(y) \Leftrightarrow w(y)=e^{-\eta y+\ln\E[e^{\eta\by}]}.
\label{eq:exptwist0}
\eeq
When $\by$ is not a continuous random variable, the exponential twisting can be rephrased in terms of probability measures $m$ and $\tilde m$ as:
\beq
\tilde m(dy)= e^{\eta y-\ln\E[e^{\eta\by}]}\, m(dy).
\eeq
The results presented in this section hold, {\em mutatis mutandis}, for this setting. 

The aforementioned change of measure was originally proposed by Cram\'er~\cite{CramerFundamental} to compute large deviations exponents, and is accordingly also known as Cram\'er's transform. 
The choice of a given exponential twisting (i.e., the choice of the parameter $\eta$) is critical in determining the accuracy of the estimates produced by the importance sampling algorithm. Interestingly, theoretical studies suggest to use for importance sampling in general exactly the {\em same} exponential twisting needed to compute the error exponents --- see, e.g.,~\cite{BucklewBook}.  
For the classical and simplest case of summation of i.i.d. random variables, the latter kind of exponential twisting is well-known --- see, e.g.,~\cite{Dembo-Zeitouni,DenHollander}. For our specific problem of adaptive distributed detection, the solution is more involved, and will be discussed in the next section. In particular, it will be shown that the theoretical results established in this paper are crucial to enable an accurate design of the importance sampling simulations.

\subsection{Importance Sampling for Adaptive Distributed Detection}
Since we are interested in evaluating the steady-state performance of the adaptive distributed algorithm, it is sufficient to examine the behavior, for $n$ sufficiently large, of the random variable $\by^\star_k(n)$ introduced in~(\ref{eq:finitehorizykn}), which, according to~(\ref{eq:ystarknnulltransient}), {\em has the same distribution of} the diffusion output with null transient.

In order to implement the importance sampling recipe, we need to perform the exponential twisting~(\ref{eq:exptwist0}). As already mentioned, the parameter $\eta$ will be chosen as that value corresponding to the change of measure used to compute the error exponents. For the specific case of adaptive distributed detection, this exponential twisting is extensively discussed and employed in Appendix B --- see, e.g.,~(\ref{eq:exptraappB}) and~(\ref{eq:exptraappB2}), and amounts to selecting $\eta=\theta_\gamma/\mu$, where $\theta_\gamma$ is the solution to the stationary equation~(\ref{eq:stateqsol}) in Theorem 3. Denoting by
\beq
\phi_{k,\mu}(t;n)\dfz\ln\E[e^{t \by^\star_k(n)}]
=\sum_{i=1}^n\sum_{\ell=1}^S \psi\left(\mu(1-\mu)^{i-1}b_{k,\ell}(i) t\right),
\label{eq:LMGFtrunco}
\eeq
the LMGF of the random variable $\by^\star_k(n)$ (the additive form above comes simply from the i.i.d. assumption on the $\{\bx_\ell(i)\}$), the choice $\eta=\theta_\gamma/\mu$ applied in~(\ref{eq:exptwist0}) then yields:
\beq
w(y)=e^{-\frac{\theta_\gamma}{\mu} \,y + \phi_{k,\mu}\left(\frac{\theta_\gamma}{\mu};n\right)
}.
\label{eq:yexptwist}
\eeq
Unfortunately, in order to generate samples according to the transformed pdf $\tilde f(y)$ in~(\ref{eq:exptwist0}), the weighting function $w(y)$ is not sufficient, since one needs to know also the pdf $f(y)$ of the random variable $\by^\star_k(n)$. As already observed, knowledge of this distribution is seldom available. To overcome this issue, one may consider running first a Monte Carlo simulation by generating random instances of the local statistics $\bx_{\ell}(i)$, and then evaluating $\by^\star_k(n)$ through~(\ref{eq:finitehorizykn}). 
However, recall that we are interested in implementing importance sampling. This means that we must find the appropriate change of measure applied to $\bx_\ell(i)$ (and not to $\by^\star_k(n)$) that would produce the desired exponential change of measure~(\ref{eq:yexptwist}) on the random variable $\by^\star_k(n)$. 
To this aim, let us denote by $p(x)$ the pdf of $\bx_\ell(i)$. We now show that the desired goal can be achieved by drawing the $(i,\ell)$-th sample $\bx_\ell(i)$ from the pdf
\beq
\tilde p_{i,\ell}(x)=e^{ \eta_{i,\ell} x - \psi(\eta_{i,\ell})} p(x),
\label{eq:piellex}
\eeq
with
\beq
\eta_{i,\ell}\dfz (1-\mu)^{i-1} b_{k,\ell}(i) \,\theta_\gamma.
\eeq
All these samples are still generated independently, but now they are no longer identically distributed.
Note also that  the above transformation depends upon the index $k$ of the agent under consideration, even if the subscript has been suppressed for ease of notation.

To see why the choice~(\ref{eq:piellex}) will correspond to~(\ref{eq:yexptwist}), let us introduce the joint ensemble $\bm{X}=\{\bx_\ell(i)\}$, for $i=1,2,\dots, n$ and for $\ell=1,2,\dots,S$.
The expectation in~(\ref{eq:basicOfIs}) can then be rewritten in terms of the distribution of $\bm X$ as follows:
\beq
\P[\by^\star_k(n)>\gamma]=\E[{\cal I}_{\{\by^\star_k(n)>\gamma\}}]
=
\E_{\tilde p}\left[\frac{p(\bm{X})}{\tilde p(\bm{X})}{\cal I}_{\{\by^\star_k(n)>\gamma\}}\right].
\label{eq:intermsofp}
\eeq 
But since the $\bx_{\ell}(i)$ are spatially and temporally i.i.d., we have from~(\ref{eq:piellex}) and~(\ref{eq:yexptwist}):
\beqa
\frac{p(\bm{X})}{\tilde p(\bm{X})}&=&e^{-\sum_{i=1}^n\sum_{\ell=1}^S \eta_{i,\ell} \bx_{\ell}(i)+\sum_{i=1}^n\sum_{\ell=1}^S \psi(\eta_{i,\ell})}\nonumber\\
&=& e^{-\frac{\theta_\gamma}{\mu}\by^\star_k(n) + \phi_{k,\mu}\left(\frac{\theta_\gamma}{\mu};n\right)}=w(\by^\star_k(n)),
\eeqa
where, in the last equality, we applied definitions~(\ref{eq:finitehorizykn}) and~(\ref{eq:LMGFtrunco}). This result shows that~(\ref{eq:intermsofp}) corresponds to the expectation in~(\ref{eq:basicOfIs}) with the pdf $\tilde f(y)$ chosen as in~(\ref{eq:yexptwist}). 

Before ending this section, we stress that the above considerations show one further important benefit: {\em our large deviations results about distributed detection over adaptive networks are also very useful in performing a careful design of a Monte Carlo simulator based on importance sampling.}

\section{Illustrative Design}
\label{sec:numexamp}

\subsection{Network Topology and Combination Weights}
We consider a network made of $S=10$ sensors, arranged so as to form the topology in the inset of Fig.~\ref{fig:fig1}.
Given the topology, two different combination matrices will be tested. The first one is defined by the so-called Metropolis rule~\cite{SayedProcIEEE}. Denoting by ${\cal N}_k$ the neighborhood of the $k$-th agent (including $k$ itself), and by $n_k$ the  cardinality $|{\cal N}_k|$ (aka the degree of the $k$-th agent), the Metropolis rule is defined by:
\beq
a_{k,\ell}=\left\{
\begin{array}{lll}
&1/\max\{n_k,n_\ell\},\qquad &\ell\in{\cal N}_k\setminus \{k\},
\\
\\
&1-\displaystyle{\sum_{m\in{\cal N}_k\setminus \{k\}}}a_{k,m},\qquad &\ell=k,
\\
&0,\qquad &\ell\notin {\cal N}_k.
\end{array}
\right.
\label{eq:Metropolis}
\eeq
This choice provides a doubly-stochastic $A$, and, hence, the corresponding Perron eigenvector has uniform entries, $p_\ell=1/S$ for all $\ell=1,2,\dots, S$.

The second combination policy is the uniform averaging rule~\cite{SayedProcIEEE}:
\beq
a_{k,\ell}=\left\{
\begin{array}{lll}
&1/n_k,\qquad &\ell\in{\cal N}_k,\\
\\
&0,\qquad &\ell\notin {\cal N}_k.
\end{array}
\right.
\label{eq:unifaverage}
\eeq
This choice provides a right-stochastic $A$, whose Perron eigenvector is available in closed form, and has entries given by~\cite{SayedProcIEEE}:
\beq
p_\ell=\frac{n_\ell}{\sum_{m=1}^S n_m},\quad \ell=1,2,\dots,S.
\eeq

\subsection{Hypothesis Test, Local Statistics and Simulation}
\label{subsec:detlaplace}
We examine the following canonical shift-in-mean detection problem with noise distributed according to a Laplace distribution, which is considered in~\cite{AdaptiveDetectionArxiv,ADD_ICASSP2014} as well.
The Laplace pdf (with scale parameter set to $1$, without loss of generality) will be denoted by:
\beq
\mathscr{L}(d)=\frac{1}{2}e^{-|d|}.
\eeq
The hypothesis test can then be formulated as follows: 
\beqa
{\cal H}_0&:&\bd_k(n)\sim \mathscr{L}(d),\\
{\cal H}_1&:&\bd_k(n)\sim\mathscr{L}(d-\rho),
\label{eq:LaplaceTest}
\eeqa
where $\bd_k(n)$ denotes the measurement collected by agent $k$ at time $n$, and $\rho>0$ is the shift-in-mean parameter. The local statistics $\bx_k(n)$ are chosen as the local log-likelihood ratios:
\beq
\bx_k(n)=\ln\left(\frac{\mathscr{L}(\bd_k(n)-\rho)}{\mathscr{L}(\bd_k(n))}\right)=|\bd_k(n)| - |\bd_k(n)-\rho|.
\label{eq:xknexpr}
\eeq 
Even if the above relationship shows clearly how to obtain the random variable $\bx_k(n)$ from the knowledge of $\bd_k(n)$, it is useful, for later use, to evaluate explicitly the distribution of $\bx_k(n)$ under the two hypotheses. 
To this aim, we rewrite~(\ref{eq:xknexpr}) as:
\beq
\bx_{k}(n)=\left\{
\begin{array}{lll}
&-\rho,\qquad & \bd_k(n)<0,
\\
&+\rho,\qquad & \bd_k(n)>\rho,
\\
&2\bd_k(n)-\rho,\qquad & \bd_k(n)\in[0,\rho],
\end{array}
\right.
\label{eq:xlimiter}
\eeq
which is obtained by using, in the three ranges  considered in~(\ref{eq:xlimiter}), the explicit definition of the absolute values appearing in~(\ref{eq:xknexpr}).
We see then that $\bx_k(n)$ is a random variable of mixed type, taking values in the range $[-\rho,\rho]$, and with two atoms located at $\pm \rho$. 
For ease of description, we find convenient to use, for random variables of mixed type, the generalized pdf written using the Dirac-delta function $\delta(x)$. 

Accordingly, let $p_0(x)$ and $p_1(x)$ denote the generalized pdfs of $\bx_k(n)$ under ${\cal H}_0$ and ${\cal H}_1$, respectively. For a shift-in-mean with respect to a symmetric pdf (as $\mathscr{L}(d)$ is), it is well-known that the log-likelihood ratio exhibits the symmetry property: $p_1(x)=p_0(-x)$ --- see, e.g.,~\cite{VanTreesBook}. Thus, it suffices to focus on $p_0(x)$. To this aim, we observe that, in view of~(\ref{eq:xlimiter}), the cumulative distribution function of $\bx_k(n)$ is 
\beq
\P_0[\bx_k(n)\leq x]=\left\{
\begin{array}{lll}
&0, & x<-\rho,
\\
\\
&\P_0\left[\bd_k(n)\leq \displaystyle{\frac{x+\rho}{2}}\right], & x\in[-\rho,\rho),
\\
\\
&1, &x\geq\rho,
\end{array}
\right.
\eeq
which corresponds to the following generalized pdf:
\beqa
\lefteqn{p_0(x)=}\nonumber\\
&&\P_0[\bd_k(n)<0]\delta(x+\rho) + \P_0[\bd_k(n)>\rho]\delta(x-\rho)+\nonumber\\
&&
\frac 1 2\mathscr{L}\left(\frac{x+\rho}{2}\right)\Pi\left(\frac{x}{2\rho}\right)\nonumber\\
&=&
\frac{1}{2}\delta(x+\rho) + \frac{e^{-\rho}}{2}\delta(x-\rho)+
\frac{1}{4}e^{-\frac{x+\rho}{2}}\Pi\left(\frac{x}{2\rho}\right),
\label{eq:p0xgenpdf}
\eeqa
where $\Pi(x)$ is a unit-width rectangular window centered at $0$. The corresponding LMGF of $\bx_k(n)$ is computable in closed form:
\beqa
\psi_0(t)&=&\ln\E_0[e^{t\bx_k(n)}]=\ln\left(\int_{-\infty}^\infty e^{t x}p_0(x)dx\right)\nonumber\\
&=&
\ln\left(
\frac{e^{-t \rho}}{2} + \frac{e^{(t-1) \rho}}{2}+
\frac{e^{-\rho/2} \rho}{2}\, \textnormal{sinch}[\rho(t-1/2)]
\right),\nonumber\\
\label{eq:psi0}
\eeqa
where
\beq
\textnormal{sinch}(x)\dfz\frac{\sinh(x)}{x},\quad
\textnormal{sinch}(0)=1.
\eeq
Since, as observed, $p_1(x)=p_0(-x)$, the LMGF under ${\cal H}_1$ is easily obtained as $\psi_1(t)=\psi_0(-t)$.
We note in passing that the above explanation and the following derivations can be restated in a more formal language by using, for the mixed-type random variable $\bx_k(n)$, a probability measure made of the superposition of two singular, atomic measures (with masses located at $\pm\rho$), and an absolutely continuous measure with density given by the third term in~(\ref{eq:p0xgenpdf}).

\vspace*{5pt}
Before concluding this section, it remains to show how to implement, for the considered Laplace example, the importance sampling method described in Sec.~\ref{sec:ISMC}.
We shall focus on hypothesis ${\cal H}_0$, and, again, the results for ${\cal H}_1$ can be simply obtained from the relationship $p_1(x)=p_0(-x)$.
Applying an exponential twisting with parameter $\eta$ to the generalized pdf in~(\ref{eq:p0xgenpdf}), we get:
\beqa
\tilde p_0(x)&=&e^{\eta x-\psi_0(\eta)}p_0(x)\nonumber\\
&=&\frac{e^{-\eta\rho}}{2e^{\psi_0(\eta)}}\delta(x+\rho) +\nonumber\\
&& \frac{e^{(\eta-1) \rho}}{2e^{\psi_0(\eta)}}\delta(x-\rho)+\nonumber\\
&&
\frac{e^{-\rho/2}}{4e^{\psi_0(\eta)}}e^{x(\eta-1/2)}\Pi\left(\frac{x}{2\rho}\right).
\label{eq:p0xtilde}
\eeqa
Introducing the definitions: 
\beq
p_{-}=\frac{e^{-\eta \rho}}{2e^{\psi_0(\eta)}},\qquad
p_{+}=\frac{e^{(\eta-1)\rho}}{2e^{\psi_0(\eta)}},
\eeq
and using~(\ref{eq:psi0}), by simple algebra relation~(\ref{eq:p0xtilde}) becomes:
\beqa
\tilde p_0(x)&=&p_{-}\delta(x+\rho) + p_{+}\delta(x-\rho) +[1-p_{-}-p_{+}]\times\nonumber\\
&& 
\frac{1}{2\rho}\,\frac{e^{x(\eta-1/2)}}{ \textnormal{sinch}[\rho (\eta-1/2)]}\Pi\left(\frac{x}{2\rho}\right).
\label{eq:finaltildep0}
\eeqa
Generating a random variable $\bx_k(n)$ according to the distribution $\tilde p_0(x)$ is now an easy task. As a matter of fact,~(\ref{eq:finaltildep0}) reveals that $ \tilde p_0(x)$ is a mixture of three components such that $\bx_k(n)$ takes on the value $-\rho$ with probability $p_{-}$, the value $+\rho$ with probability $p_{+}$, and that otherwise it must be sampled from the pdf
\beq
\frac{1}{2\rho}\, \frac{e^{x(\eta-1/2)}}{\textnormal{sinch}[\rho(\eta-1/2)]}\Pi\left(\frac{x}{2\rho}\right).
\eeq

\subsection{A Normal Approximation}
In the following analysis, we shall compare the error probabilities estimated empirically, to the refined asymptotic formulas obtained in the present work. As a further term of comparison, we would like to add a normal approximation that will follow from the asymptotic normality result proved in~\cite{AdaptiveDetectionArxiv,ADD_ICASSP2014}. Actually, this result was obtained there for doubly-stochastic connection matrices, but the generalization to the case of right-stochastic matrices comes essentially at no cost and can be stated as:
\beq
\boxed
{
\frac{\by^\star_{k,\mu}-\E[\bx]}{\sqrt{\mu\,\sigma^2_{\lim}}}\stackrel{\mu\rightarrow 0}{\rightsquigarrow} {\cal N}(0,1)
}
\label{eq:CLT}
\eeq
where the limiting variance $\sigma^2_{\lim}$ can be obtained by applying result~(\ref{eq:Theorem1_5}) to the second cumulant (i.e., the variance) of $\by^\star_{k,\mu}$, which yields:
\beq
\frac{\VAR[\by^\star_{k,\mu}]}{\mu}\stackrel{\mu\rightarrow 0}{\longrightarrow} \frac{\sigma^2_x}{2}\sum_{\ell=1}^S p^2_\ell \dfz \sigma^2_{\lim}.
\label{eq:sigma2lim}
\eeq
We emphasize that this result is consistent with previous findings obtained in the context of mean-square-error estimation --- see, e.g.,~\cite{SayedSPmag}.

We see from~(\ref{eq:sigma2lim}) that the ratio $\VAR[\by^\star_{k,\mu}]/(\mu\sigma^2_{\lim})$ converges to one as $\mu$ goes to zero. In view of Slutsky's Theorem~\cite{shao}, this fact implies that the following alternative version of the convergence in distribution in~(\ref{eq:CLT}) holds:
\beq
\boxed
{
\frac{\by^\star_{k,\mu}-\E[\bx]}{\sqrt{\VAR[\by^\star_{k,\mu}]}}\stackrel{\mu\rightarrow 0}{\rightsquigarrow} {\cal N}(0,1)
}
\label{eq:CLT2}
\eeq
While the two formulations~(\ref{eq:CLT}) and~(\ref{eq:CLT2}) are asymptotically equivalent, it is expected that~(\ref{eq:CLT2}) offers a better performance since it replaces the asymptotic variance with the actual one. The relationship shown in~(\ref{eq:CLT2}) suggests the  following way to approximate the probability $\P[\by^\star_{k,\mu}>\gamma]$:
\beq
\P[\by^\star_{k,\mu}>\gamma]\approx
Q\left(\frac{\gamma-\E[\bx]}{\sqrt{\VAR[\by^\star_{k,\mu}]}}\right),
\label{eq:normapp}
\eeq
where $Q(\cdot)$ denotes the complementary cumulative distribution function of a standard normal distribution.

\subsection{Analysis of the Results}

Let us now examine the adaptive distributed network of detectors in operation, by reporting the evidence arising from our Monte Carlo analysis. 
We refer to a sufficiently large time horizon, such that the steady-state  assumption applies, and evaluate the error probabilities for different values of the step-size. 

The exact asymptotics provided by Theorem 3 will be computed by implementing the six-steps recipe described at the end of Sec.~\ref{sec:maintheorems}. Clearly, in doing so, we must use the LMGF $\psi_0(t)$ if we are working under ${\cal H}_0$, and the LMGF $\psi_1(t)$ if we are working under ${\cal H}_1$. 
The normal approximation will be instead obtained as described in the previous section --- see~(\ref{eq:normapp}).

\vspace*{5pt}
In the two examples that we are going to discuss we choose the detection threshold as detailed in~\cite{AdaptiveDetectionArxiv,ADD_ICASSP2014}, obtaining $\gamma=0$.
This implies, by the symmetry property $p_1(x)=p_0(-x)$, that the error probabilities of first and second kind defined by~(\ref{eq:alphabet3}) are equal, namely:
\beq
\alpha_{k,\mu}=\beta_{k,\mu}.
\eeq
Consistently, in the following description, the terminologies ``error probability" and ``error exponent" refer to any of these errors.

\begin{figure}[t]
\centerline{\includegraphics[width=.5\textheight]{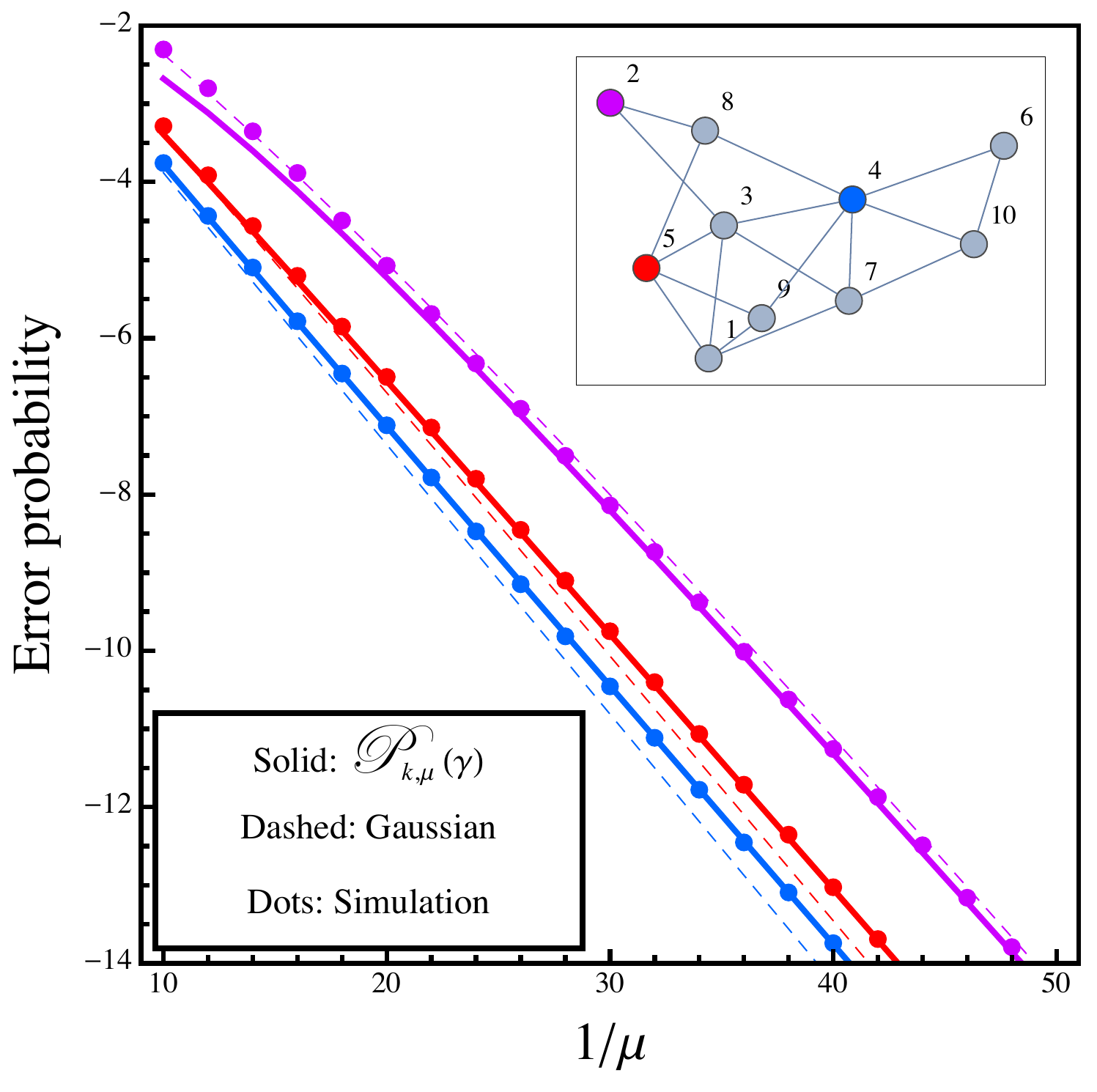}}
\caption{
Laplace example discussed in Sec.~\ref{subsec:detlaplace}, with $\rho=0.6$. 
The network topology is depicted in the inset plot, and the combination weights $a_{k,\ell}$ follow the Metropolis rule~(\ref{eq:Metropolis}). The performance of agents 2, 4 and 5 is displayed.  
Dots refer to the empirical steady-state error probabilities $\alpha_{k,\mu}=\beta_{k,\mu}$ at different sensors, obtained via Monte Carlo simulation with importance sampling, as described in~\ref{sec:ISMC}. Solid curves refer to the exact asymptotics provided by Theorem 3. Dashed curves refer to the normal approximation~(\ref{eq:normapp}). 
}
\label{fig:fig1}
\end{figure}

\begin{figure}[t]
\centerline{\includegraphics[width=.5\textheight]{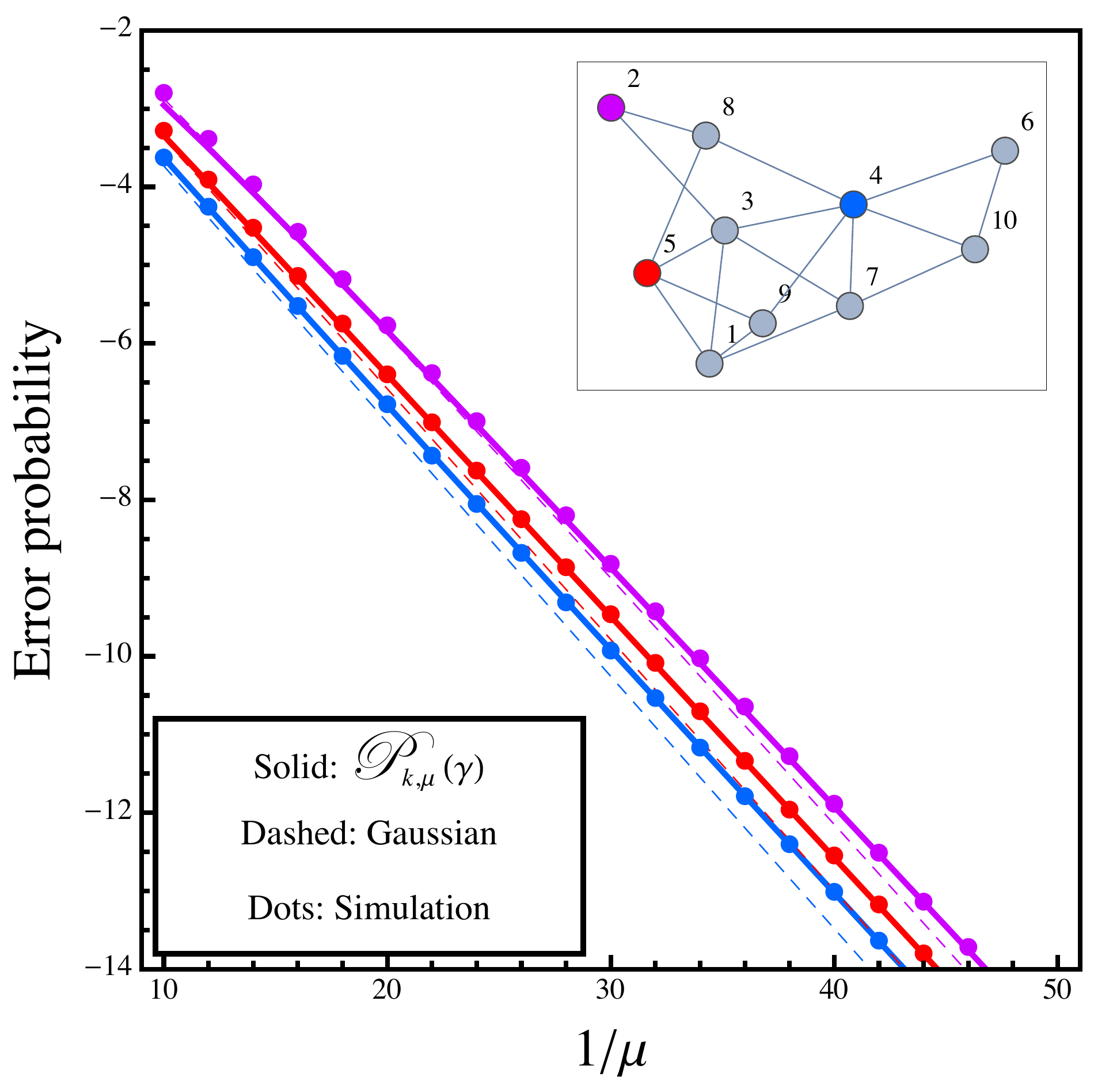}}
\caption{
Laplace example discussed in Sec.~\ref{subsec:detlaplace}, with $\rho=0.6$. 
The network topology is depicted in the inset plot, and the combination weights $a_{k,\ell}$ follow the averaging rule~(\ref{eq:unifaverage}). The performance of agents 2, 4 and 5 is displayed. Dots refer to the empirical steady-state error probabilities $\alpha_{k,\mu}=\beta_{k,\mu}$ at different sensors, obtained via Monte Carlo simulation with importance sampling, as described in~\ref{sec:ISMC}. Solid curves refer to the exact asymptotics provided by Theorem 3. Dashed curves refer to the normal approximation~(\ref{eq:normapp}). 
}
\label{fig:fig2}
\end{figure}

We start by considering the doubly-stochastic combination matrix obtained with the Metropolis rule~(\ref{eq:Metropolis}).
In Fig.~\ref{fig:fig1}, the performance of the agents is displayed as a function of $1/\mu$, and different agents are marked with different colors. 
The main system features, which were already commented in~\cite{AdaptiveDetectionArxiv,ADD_ICASSP2014}, are here summarized. 

The first evidence is that the different curves pertaining to different agents stay nearly parallel for sufficiently small values of the step-size $\mu$, i.e., the detection error probabilities at different sensors vanish exponentially as functions of $1/\mu$, sharing the same detection error exponent.
This was the main result revealed by the large deviations analysis performed in~\cite{AdaptiveDetectionArxiv,ADD_ICASSP2014}.

However, as already noticed in~\cite{AdaptiveDetectionArxiv,ADD_ICASSP2014}, the large deviations tool is not powerful enough to capture an important feature of the distributed behavior.
Indeed, the second evidence emerging from the simulations is that the error probability curves in Fig.~\ref{fig:fig1} are basically ordered, and the ordering is closely related to the network connection structure. Comparing the detection performance of three specific agents, namely, agents $2,4,5$, we see that the ordering reflects the degree of connectivity of each agent. For instance, agent $4$ has the highest number of neighbors, and its performance is the best one, while agent $2$ is the most isolated, and its error probability curve appears consistently as the highest one. According to what one expects, agent $5$ is in an intermediate position.
{\em The new fact here is that, using the results of the current manuscript, we are now able to provide a systematic analysis of the above features, as well as of the exact interplay with network connectivity.}

\vspace*{5pt}
First of all, what in~\cite{AdaptiveDetectionArxiv,ADD_ICASSP2014} was only a partial evidence arising from a particular numerical experiment, emerges now, thanks to the refined asymptotic analysis, as the {\em universal behavior of adaptive distributed detection}.

Moreover, the refined asymptotic approximations provided by Theorem 3 can be used to obtain quantitative predictions of the actual system performance, as we proceed to explain.
The refined formulas are represented by the solid curves in Fig.~\ref{fig:fig1}. We see that the empirical probability points (the dots) converge toward the theoretical solid curves as the step-size $\mu$ decreases (i.e., as we move to the right in the plot). 
Remarkably, the theoretical formulas provided by our theorems are able to embody the dependencies between the network connection structure and the detection performance at different sensors, as it is witnessed by the correct ordering of the curves. 
In addition, a gap between the exact asymptotics (solid curves) and the normal approximation (dashed curves) is clearly observed. 
This should come as no surprise, since the normal approximation is expected to be accurate when working with {\em small} deviations, and must accordingly provide a wrong prediction in the large deviations regime --- see also the discussion in~\cite{AdaptiveDetectionArxiv,ADD_ICASSP2014}.

\vspace*{5pt}
We now switch to the analysis of the right-stochastic combination matrix provided by the uniform averaging rule~(\ref{eq:unifaverage}).
The corresponding results are reported in Fig.~\ref{fig:fig2}. First, we note that, as in the doubly-stochastic case, the error probability curves vanish exponentially fast as functions of $1/\mu$, and stay nearly parallel as $\mu$ goes to zero. This is consistent with the prediction that they share the same error exponent, as dictated by Theorem 2 for the general case of {\em right}-stochastic combination matrices.
Also in this example, we are able to appreciate the goodness of the refined approximations obtained with Theorem 3, and the fact that the empirical points depart from the normal approximation.

\vspace*{5pt}
To get further insights, we now compare the detection performance of the two aforementioned combination matrices. 
To begin with, we compute the error exponents pertaining to the two systems. For the particular example considered, we obtain (recall that the threshold is set to $\gamma=0$):
\beq
\Phi^{\textnormal{DS}}(0)\approx 0.75
>
\Phi^{\textnormal{RS}}(0)\approx 0.7,
\label{eq:rightdoublyexpo}
\eeq
for the doubly-stochastic and the right-stochastic case, respectively.
These values appear to suggest that that the doubly-stochastic combination policy asymptotically outperforms the right-stochastic combination policy. This conclusion may somehow be expected because, asymptotically, a doubly-stochastic combination policy weights the local statistics equally, while a right-stochastic combination policy does not. In the presence of i.i.d. observations, the former strategy seems to be preferable.
To see if this first-order analysis suffices, let us now apply the refined asymptotic formulas.

\begin{figure}[t]
\centerline{\includegraphics[width=.5\textheight]{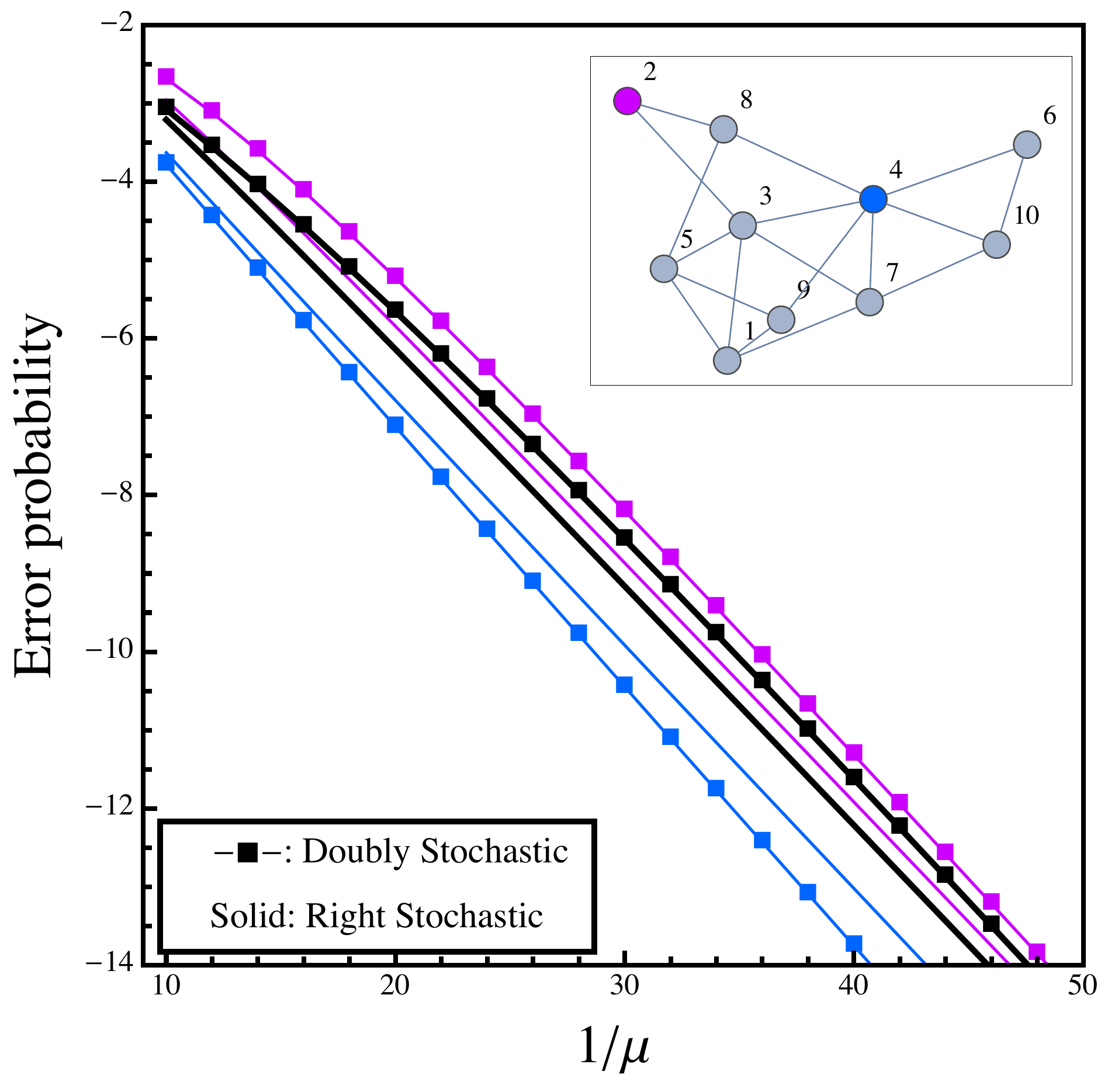}}
\caption{
Laplace example discussed in Sec.~\ref{subsec:detlaplace}, with $\rho=0.6$. 
The network topology is depicted in the inset plot, and the performance corresponding to two combination matrices is compared, namely, the doubly-stochastic matrix obtained with the Metropolis rule~(\ref{eq:Metropolis}), and the right-stochastic matrix obtained with the uniform averaging rule~(\ref{eq:unifaverage}).  
The performance of agents 2 and 4 is displayed, along with the average network performance, namely, the arithmetic mean of the error probabilities of all agents (black). All curves are computed by using the exact asymptotics provided by Theorem 3.  
}
\label{fig:fig3}
\end{figure}

Accordingly, in Fig.~\ref{fig:fig3} we display the theoretical error probability curves obtained with Theorem 3, for the Metropolis combination matrix (doubly-stochastic case, line and markers) and for the uniform averaging combination matrix (right-stochastic case, solid curves). An interesting behavior arises. 
{\em We observe that the relative performance of the two different combination policies depends strongly on the connectivity of the individual agent.} For instance, for the well-connected agent $4$, the doubly-stochastic combination policy delivers superior performance, while exactly the converse is true for the scarcely connected agent $2$. 
Moreover, if we consider as network performance the arithmetic average of the error probabilities (black curves), we see that the right-stochastic combination policy is globally superior. 

An explanation for this behavior is as follows. Denoting by $\lambda_2^{\textnormal{DS}}$ and $\lambda_2^{\textnormal{RS}}$ the second largest magnitude eigenvalues of the doubly-stochastic and right-stochastic combination matrices, respectively, we see that 
\beq
\lambda_{2}^{\textnormal{DS}}\approx  0.83> \lambda_{2}^{\textnormal{RS}}\approx 0.7,
\eeq
implying that the right-stochastic weights will converge to the corresponding Perron eigenvector faster than the doubly-stochastic weights. Examining the detection performance, we see from here that the slower convergence of the doubly-stochastic combination matrix has a detrimental effect on the less connected agents.
We could say that the benefits of the higher (doubly-stochastic case) exponent are more than compensated by the faster (right-stochastic case) convergence to the steady-state behavior.

Notice that the observed behavior is not in contrast with what is predicted by~(\ref{eq:rightdoublyexpo}). Based on an analysis at the first leading order in the exponent, the doubly-stochastic combination policy are asymptotically superior to the right-stochastic one. This means that the two curves corresponding to sensor $2$ in the figure must cross for a certain vanishingly small $\mu$. What the refined analysis is able to tell is that this value might be too small for a given regime of analysis. 

Before concluding, we would like to stress that the above comparison between combination policies should be considered preliminary, in that: $i)$ the comparison has been made for two systems operating with the same value of the step-size $\mu$, and, for a more complete view, the analysis should be complemented by examining also the transient behavior of the two combination policies; $ii)$ the evidences are obtained with reference to a particular doubly-stochastic matrix and a particular right-stochastic matrix. 
This notwithstanding, while no general conclusions can be drawn at this stage regarding the relative advantages of the two kinds of combination strategies for individual agents, a general trend seems to emerge: {\em connectivity matters}.  Since the connectivity features are embodied in the higher-order corrections, the simplest large deviations analysis is not sufficient, and the refined exact asymptotics provided by Theorem 3 are crucial in assessing the performance of adaptive distributed detection over networks.

\appendices

\vspace*{10pt}

\section{}

In the following, the $r$-th derivative of a function $f(t)$ will be denoted by $f^{(r)}(t)$, with the convention that $f^{(0)}(t)=f(t)$. When convenient, the first three derivatives will be alternatively denoted by $f^{\prime}(t)$, $f^{\prime\prime}(t)$, and $f^{\prime\prime\prime}(t)$. Moreover, the notation $f_\mu={\cal O}(\mu)$ means that the ratio $f_\mu/\mu$ stays bounded as $\mu\rightarrow 0$.

\vspace*{5pt}
\noindent
{\bf \textsc{Theorem 1} (Fundamental properties  of $\bm{\phi^{(r)}_{k,\mu}(t)}$).} 
{\em 
Assume that $\psi(t)<\infty$ for all $t\in\mathbb{R}$, and introduce the quantity: 
\beq
\xi_{i,\ell}\dfz\mu (1-\mu)^{i-1}b_{k,\ell}(i).
\label{eq:csidef}
\eeq
Then, the following facts hold:
\begin{itemize}
\item[$i)$]
The LMGF of $\by^\star_{k,\mu}$ can be computed as:
\beq
\boxed{
\phi_{k,\mu}(t)=\sum_{i=1}^\infty\sum_{\ell=1}^S \psi(\xi_{i,\ell} t)
}
\label{eq:Theorem1_1}
\eeq
and the limiting LMGF $\phi(t)$ is:
\beq
\boxed
{
\phi(t)=\sum_{\ell=1}^S 
\omega(p_\ell t)
}
\label{eq:Theorem1_2}
\eeq
\item[$ii)$]
For $r=1,2,\dots$, it holds that:
\beq
\boxed{
\phi^{(r)}_{k,\mu}(t)=\sum_{i=1}^\infty\sum_{\ell=1}^S
\xi_{i,\ell}^r \psi^{(r)}(\xi_{i,\ell} t)
}
\label{eq:Theorem1_3}
\eeq
and
\beq
\boxed{
\lim_{\mu\rightarrow 0}\frac{\phi^{(r)}_{k,\mu}(t/\mu)}{\mu^{r-1}}=\phi^{(r)}(t)
}
\label{eq:Theorem1_4}
\eeq
In particular, the $r$-th cumulant, $\phi^{(r)}_{k,\mu}(0)$, of the steady-state random variable $\by^\star_{k,\mu}$ satisfies:
\beq
\boxed{
\lim_{\mu\rightarrow 0}\frac{\phi^{(r)}_{k,\mu}(0)}{\mu^{r-1}}=\phi^{(r)}(0)=\frac{\psi^{(r)}(0)}{r}\,\sum_{\ell=1}^S p_\ell^r
}
\label{eq:Theorem1_5}
\eeq
where $\psi^{(r)}(0)$ is the $r$-th cumulant of the local statistic $\bx_k(n)$.
\item[$iii)$] With reference to the convergence in~(\ref{eq:Theorem1_2}) and~(\ref{eq:Theorem1_4}), the following refined estimate of the convergence error holds for $r=0,1,\dots$:
\beq
\frac{\phi_{k,\mu}^{(r)}(t/\mu)}{\mu^{r-1}}=\phi^{(r)}(t)+{\cal O}(\mu).
\label{eq:Theorem1_6}
\eeq
\end{itemize}
}
~\hfill$\square$

\vspace*{10pt}
\noindent
In order to establish the validity of Theorem 1, we start by proving a couple of useful lemmas.
Our first lemma  is a simple generalization of Lemmas 1 and 2 from~\cite{AdaptiveDetectionArxiv,ADD_ICASSP2014}.
We shall focus on a function $f(t)$ twice differentiable in $\mathbb{R}$, with $f(0)=0$. For such a function, we have that:
\begin{itemize}
\item
The function $f(t)/t$ is continuous for all $t\in\mathbb{R}$. For any $t\neq 0$, the result follows from the continuity of $f(t)$. For $t=0$, the result is easily verified by recalling that $f(0)=0$, yielding:
\beq
\lim_{t\rightarrow 0} \frac{f(t)}{t}=f^\prime(0).
\eeq
\item
The derivative $(d/dt)(f(t)/t)$ is continuous for all $t\in\mathbb{R}$. For any $t\neq 0$, this result follows immediately from the assumed smoothness properties of $f(t)$. For $t=0$, the result is easily verified by recalling that $f(0)=0$, and observing that 
\beq
\lim_{t\rightarrow 0} \frac{d}{dt}\frac{f(t)}{t}=\lim_{t\rightarrow 0} \frac{f^\prime(t)t-f(t)}{t^2}=\frac{f^{\prime\prime}(0)}{2},
\eeq
where we used L'Hospital's rule~\cite{RudinBook}.
\end{itemize}
We introduce the auxiliary functions:
\beq
h_1(t)=
\frac{t^2}{2} \times 
\left\{
\begin{array}{l}
\displaystyle
{
\max_{\tau\in[0, t]} \left|\frac{d}{d\tau}\frac{f(\tau)}{\tau}\right|, \quad t\geq 0,
} \\
\\
\displaystyle{
\max_{\tau\in[t, 0]} \left|\frac{d}{d\tau}\frac{f(\tau)}{\tau}\right|, \quad t<0.
}
\end{array}
\right.
\label{eq:h1def}
\eeq
and
\beq
h_2(t)=
|t| \times 
\left\{
\begin{array}{l}
\displaystyle{\max_{\tau\in[0, t]} |f^\prime(\tau)|, \quad t\geq 0,} \\
\displaystyle{\max_{\tau\in[t, 0]} |f^\prime(\tau)|, \quad t<0.}
\end{array}
\right.
\label{eq:h2def}
\eeq
We can easily show that:
\beq
0\leq h_1(t)<\infty,\qquad 0\leq h_2(t)<\infty,\qquad \forall t\in\mathbb{R}.
\eeq 
Indeed, $h_1(t)\geq 0$ and $h_2(t)\geq 0$ by definition. Finiteness of both functions follows from Weierstrass extreme value theorem~\cite{RudinBook} since, by the properties for $f(t)$ and $f(t)/t$ discussed above, the maxima appearing in~(\ref{eq:h1def}) and~(\ref{eq:h2def}) are maxima of continuous functions over compact sets for any finite $t$.

\vspace*{5pt}
\noindent
{\bf \textsc{Lemma 1}.}
{\em
Let $f(t)$ be twice continuously differentiable in $\mathbb{R}$, with $f(0)=0$. Let, for all $i\in\mathbb{N}$: 
\beq
0\leq b_i\leq 1,\quad b_i\stackrel{i\rightarrow\infty}{\longrightarrow} p>0,\qquad \textnormal{with }
\left |b_i-p\right |\leq {\cal C}\lambda^i
\label{eq:Perronlikeassump}
\eeq
for some ${\cal C}>0$ and $0<\lambda<1$.
Then, we have, for all $t\in\mathbb{R}$:
\beq
\boxed{
\sum_{i=1}^\infty f(\mu (1-\mu)^{i-1} b_i t)
=
\frac 1 \mu \int_{0}^{\mu p t}\frac{f(\tau)}{\tau}d\tau + R(t)
}
\label{eq:mainlemmaclaim}
\eeq
where:
\beq
\boxed{
|R(t)|\leq
\frac{h_1(\mu p t)}{2-\mu}+
\frac{{\cal C} \lambda h_2(\mu t)}{1-\lambda(1-\mu)}
}
\label{eq:Restimate}
\eeq
}

\vspace*{5pt}
\noindent
{\em Proof}. First, note that the integral in~(\ref{eq:mainlemmaclaim}) is well-posed, since the function $f(t)/t$ is continuous for all $t\in\mathbb{R}$. Then, for the case $t=0$, Eq.~(\ref{eq:mainlemmaclaim}) is trivially verified, since $f(0)=0$ by assumption, and $h_1(0)=h_2(0)=0$ by definitions~(\ref{eq:h1def}) and~(\ref{eq:h2def}).  
We now prove the lemma for the case $t>0$, and the proof for $t<0$ will follow similarly. 

Let us introduce the infinite partition of the interval $[0,\mu p t]$:
\beq
\tau_i=\mu (1-\mu)^{i-1} p t,\qquad
i=1,2,\dots
\eeq
By using a first-order Taylor expansion we can write:
\beqa
\lefteqn{
\sum_{i=1}^n
f(\mu (1-\mu)^{i-1} b_i t)
}\nonumber\\
&=&
\sum_{i=1}^n
f(\tau_i) + 
\sum_{i=1}^n 
\mu (1-\mu)^{i-1}
f^{\prime}(t_i)
(b_i-p) t.\nonumber\\
\label{eq:mazingaz1}
\eeqa
for a value $t_i$ that is certainly contained in the interval $[0,\mu t]$, because so are the points $\mu(1-\mu)^{i-1} b_i t$ and $\tau_i$. Let us focus on the first term on the RHS of~(\ref{eq:mazingaz1}), and let $g(t)=f(t)/t$. A second-order Taylor expansion of the function $G(t)=\int_{t}^{\tau_i} g(\tau)d\tau$ around the point $\tau_i$ gives:
\beqa
G(\tau_{i+1})&=&G(\tau_i)+g(\tau_i)\delta_i
-g^\prime(\bar t_i)\frac{\delta_i^2}{2}\nonumber\\
&=&
g(\tau_i)\delta_i
-g^\prime(\bar t_i)\frac{\delta_i^2}{2},
\label{eq:GforTaylor}
\eeqa
for a certain $\bar{t}_i\in(\tau_{i+1},\tau_i)$ and where
\beq
\delta_i=\tau_i-\tau_{i+1}=\mu\,\tau_i.
\eeq
We then obtain from~(\ref{eq:GforTaylor}) that
\beqa
\int_{\mu(1-\mu)^n p t}^{\mu p t} g(\tau) d\tau
&=&
\sum_{i=1}^n
\int_{\tau_{i+1}}^{\tau_{i}} g(\tau) d\tau
\nonumber\\
&=&
\sum_{i=1}^n
g(\tau_i)\delta_i
-
\sum_{i=1}^n
g^\prime(\bar t_i)\frac{\delta_i^2}{2}\nonumber\\
&=&
\mu \sum_{i=1}^n
f(\tau_i)
-
\sum_{i=1}^n
g^\prime(\bar t_i)\frac{\delta_i^2}{2},\nonumber\\
\label{eq:repeatedmeanval}
\eeqa
Computing now $\sum_{i=1}^n f(\tau_i)$ from~(\ref{eq:repeatedmeanval}), and substituting into~(\ref{eq:mazingaz1}), we have
\beq
\sum_{i=1}^n
f(\mu (1-\mu)^{i-1} b_i t)=
\frac{1}{\mu}\int_{\mu(1-\mu)^n p t}^{\mu p t} \frac{f(\tau)}{\tau} d\tau
+r_n(t),
\label{eq:Snfirst}
\eeq
where
\beqa
r_n(t)&\dfz&
\frac 1 \mu\sum_{i=1}^n
g^\prime(\bar t_i)\frac{\delta_i^2}{2}+\nonumber\\
&&
\sum_{i=1}^n 
\mu (1-\mu)^{i-1}
f^{\prime}(t_i)
(b_i-p) t.
\label{eq:remainder1}
\eeqa
Both series on the RHS of~(\ref{eq:remainder1}) are absolutely convergent as $n\rightarrow\infty$. Indeed, using the definition of the auxiliary function $h_1(t)$ in~(\ref{eq:h1def}), we see that the first series satisfies:
\beqa
\frac 1 \mu\sum_{i=1}^n
|g^\prime(\bar t_i)|\frac{\delta_i^2}{2}
&=&
\frac{\mu}{2}
(\mu p t)^2\sum_{i=1}^n |g^\prime(\bar t_i)| (1-\mu)^{2(i-1)}\nonumber\\
&\leq&
\mu\sum_{i=1}^\infty (1-\mu)^{2(i-1)}h_1(\mu p t)
=
\frac{h_1(\mu p t)}{2-\mu}.\nonumber\\
\eeqa
Similarly, using the definition of $h_2(t)$ in~(\ref{eq:h2def}), and the assumption $\left |b_i-p\right |\leq {\cal C}\lambda^i$, we see that the second series obeys:
\beqa
\lefteqn{
\sum_{i=1}^n 
\mu (1-\mu)^{i-1}
|f^{\prime}(t_i)
(b_i-p)| t}\nonumber\\
&\leq&
h_2(\mu t)\sum_{i=1}^n (1-\mu)^{i-1}|b_i-p|
\nonumber\\
&\leq&
{\cal C} \lambda h_2(\mu t)\sum_{i=1}^\infty [\lambda (1-\mu)]^{i-1}
\nonumber\\
&=&
\frac{{\cal C} \lambda h_2(\mu t)}{1-\lambda(1-\mu)}.
\eeqa
Convergence of the first term on the RHS in~(\ref{eq:Snfirst}) follows by the definition of integration. If we now denote by $R(t)$ the limit of $r_n(t)$ as $n\rightarrow\infty$, we have in fact proved~(\ref{eq:mainlemmaclaim}) along with the upper bound in~(\ref{eq:Restimate}).

~\hfill$\square$

\vspace*{10pt}
\noindent 
The next lemma establishes some useful properties of the function $\omega(t)$ defined in~(\ref{eq:omegadef}).

\vspace*{5pt}
\noindent
{\bf \textsc{Lemma 2} (Derivatives of $\bm{\omega(t)}$).}
{\em
Let $\psi(t)$ be the LMGF of the local statistic $\bx_k(n)$, and assume that $\psi(t)<\infty$ for all $t\in\mathbb{R}$. Then, the function $\omega(t)$ in~(\ref{eq:omegadef}) is infinitely differentiable in $\mathbb{R}$, and, for all $r\in\mathbb{N}$:
\beq
\boxed
{
\omega^{(r)}(t)=\frac{1}{t^r}\int_0^{t}\psi^{(r)}(\tau)\tau^{r-1}d\tau
}
\label{eq:omegaderiv}
\eeq
where, for $t=0$, the above equation should be read as:
\beq
\omega^{(r)}(0)=\lim_{t\rightarrow 0}\frac{1}{t^r}\int_0^{t}\psi^{(r)}(\tau)\tau^{r-1}d\tau=\frac{\psi^{(r)}(0)}{r}.
\label{eq:omega0derivo}
\eeq
}

\vspace*{5pt}
\noindent
{\em Proof}. 
We prove the lemma by induction. Let us consider first the case $t\neq 0$.
Property~(\ref{eq:omegaderiv}) holds for $r=1$:
\beq
\omega^{\prime}(t)=\frac{\psi(t)}{t}
=\frac{1}{t}\int_0^{t}\psi^{\prime}(\tau)d\tau,
\eeq
having used $\psi(0)=0$ since $\psi(t)$ is a LMGF.
Now we show that if~(\ref{eq:omegaderiv}) holds for some $r$, then it must hold for $r+1$.
Indeed, note that
\beq
\omega^{(r+1)}(t)=\frac{d}{dt}\omega^{(r)}(t)=
\frac{d}{dt}
\left(\frac{1}{t^r}\int_0^{t}\psi^{(r)}(\tau)\tau^{r-1}d\tau\right).
\eeq
Applying the rule of integration by parts we have:
\beqa
\lefteqn{\frac{1}{t^r}\int_0^{t}\psi^{(r)}(\tau)\tau^{r-1}d\tau}
\nonumber\\
&=&
\frac{1}{t^r}\left[\psi^{(r)}(\tau)\frac{\tau^r}{r}\right]_{0}^{t}
-
\frac{1}{t^r}\int_0^{t}\psi^{(r+1)}(\tau)\frac{\tau^{r}}{r}d\tau
\nonumber\\
&=&
\frac{\psi^{(r)}(t)}{r}
-
\frac{1}{r t^r}\int_0^{t}\psi^{(r+1)}(\tau)\tau^{r}d\tau.
\eeqa
Differentiating the above expression yields:
\beqa
\omega^{(r+1)}(t)&=&
\frac{\psi^{(r+1)}(t)}{r}
-
\frac{\psi^{(r+1)}(t)}{r}\nonumber\\
&+&\frac{1}{t^{r+1}}\int_{0}^{t}\psi^{(r+1)}(\tau) \tau^r d\tau.
\eeqa
This proves the claim for $t\neq 0$. 
Now, since $\omega^{(r)}(t)$ exists for all $t\neq 0$, we have~\cite{RudinBook}:
\beq
\omega^{(r)}(0)=\lim_{t\rightarrow 0}\omega^{(r)}(t),
\eeq
provided that the above limit exists. By applying L'Hospital's rule~\cite{RudinBook} we have in fact:
\beq
\omega^{(r)}(0)=\lim_{t\rightarrow 0} \frac{1}{t^{r}}\int_{0}^{t}\psi^{(r)}(\tau) \tau^{r-1} d\tau=\frac{\psi^{(r)}(0)}{r}
\eeq
and the proof is complete.

~\hfill$\square$

\vspace*{10pt}
\noindent
We are now ready to prove Theorem 1.

\vspace*{10pt}
\noindent
{\em Proof of Theorem 1}.
Recalling the definition of $\xi_{i,\ell}$ in~(\ref{eq:csidef}), the finite-horizon random variable $\by_k^{\star}(n)$ in~(\ref{eq:finitehorizykn}), and its LMGF $\phi_{k,\mu}(t;n)$ in~(\ref{eq:LMGFtrunco}) can be written as, respectively:
\beq
\by_k^{\star}(n)\dfz 
\sum_{i=1}^n \sum_{\ell=1}^S \xi_{i,\ell}\bx_{\ell}(i),
\label{eq:regimeterm}
\eeq
and
\beq
\phi_{k,\mu}(t;n)=\sum_{i=1}^n\sum_{\ell=1}^S \psi(\xi_{i,\ell}t)=\sum_{\ell=1}^S \sum_{i=1}^n \psi(\xi_{i,\ell}t).
\label{eq:phikmutn}
\eeq
We now apply Lemma 1 to each of the $S$ inner summations, with the choices $f(t)=\psi(t)$, $b_i=b_{k,\ell}(i)$, and $p=p_\ell$. Note that the hypotheses of the lemma are satisfied. Indeed $\psi(t)$ is a LMGF, implying that $\psi(0)=0$ and that $\psi(t)$ is infinitely differentiable in $\mathbb{R}$. By assumption, we have $\psi(t)<\infty$ for all $t\in\mathbb{R}$. As to the conditions in~(\ref{eq:Perronlikeassump}), by Perron's theorem~\cite[Theorem 8.5.1]{Johnson-Horn}, there exist ${\cal C}>0$ and $0<\lambda<1$ such that:
\beq
\left |b_{k,\ell}(i)-p_{\ell}\right |\leq {\cal C}\lambda^i.
\label{eq:Perron}
\eeq
By Lemma 1 we conclude that 
\beqa
\lim_{n\rightarrow\infty} \phi_{k,\mu}(t;n)&=&
\sum_{i=1}^\infty\sum_{\ell=1}^S \psi(\xi_{i,\ell}t)
\nonumber\\
&=&
\frac{1}{\mu}
\sum_{\ell=1}^S \int_{0}^{p_\ell\mu t}\frac{\psi(\tau)}{\tau}d\tau +
\sum_{\ell=1}^S R_\ell(t),\nonumber\\
\label{eq:phikmuintermediate0}
\eeqa
where we made explicit the dependence of the remainder terms upon $\ell=1,2,\dots,S$. 

In view of the continuity theorem for the moment generating functions~\cite{CurtissMGF}, and since by definition $\by^\star_k(n)$ converges in distribution to $\by^\star_{k,\mu}$, the limit of $\phi_{k,\mu}(t;n)$ (i.e., the limit of the LMGF of $\by^\star_k(n)$) represents $\phi_{k,\mu}(t)$ (i.e., the LMGF of $\by^\star_{k,\mu}$). Formally, we have:
\beq
\phi_{k,\mu}(t)=\lim_{n\rightarrow\infty} \phi_{k,\mu}(t;n)=
\sum_{i=1}^\infty\sum_{\ell=1}^S \psi(\xi_{i,\ell}t),
\eeq
and claim~(\ref{eq:Theorem1_1}) is proved. For the same reason we can rewrite~(\ref{eq:phikmuintermediate0}) as: 
\beq
\phi_{k,\mu}(t)=\frac{1}{\mu}
\sum_{\ell=1}^S \int_{0}^{p_\ell\mu t}\frac{\psi(\tau)}{\tau}d\tau + \sum_{\ell=1}^S R_\ell(t),
\label{eq:phikmuintermediate}
\eeq 
and
\beq
\mu\,\phi_{k,\mu}(t/\mu)=
\sum_{\ell=1}^S \int_{0}^{p_\ell t}\frac{\psi(\tau)}{\tau}d\tau + \mu \sum_{\ell=1}^S R_\ell(t/\mu),
\eeq 
where, from~(\ref{eq:Restimate}), the remainder term is bounded by:
\beq
\mu \sum_{\ell=1}^S |R_\ell(t/\mu)|\leq
\frac{\mu}{2-\mu}\,\sum_{\ell=1}^S h_1(p_\ell t) +
\frac{\mu\,S {\cal C} \lambda h_2(t)}{1-\lambda(1-\mu)}.
\eeq
We conclude from~(\ref{eq:phikmuintermediate}) that claim~(\ref{eq:Theorem1_2}) and claim~(\ref{eq:Theorem1_6}) for the case $r=0$  hold.

To establish the remaining claims, we start by noting from~(\ref{eq:phikmutn}) that, for all $r=1,2,\dots$:
\beq
\phi^{(r)}_{k,\mu}(t;n)=\sum_{i=1}^n\sum_{\ell=1}^S \xi_{i,\ell}^r \psi^{(r)}(\xi_{i,\ell}t).
\label{eq:phiderivativestn}
\eeq
Now, recalling the definition of $\xi_{i,\ell}$ from~(\ref{eq:csidef}), we have:
\beq
0\leq \xi_{i,\ell}\leq \mu(1-\mu)^{i-1},\qquad
t\in[-a,a] \Rightarrow \xi_{i,\ell} t\in [-a,a].
\eeq
Thus, for all $t\in[-a,a]$ we can write:
\beq
|\xi_{i,\ell}^r \psi^{(r)}(\xi_{i,\ell}t)|\leq
\mu^r (1-\mu)^{r(i-1)} \max_{\tau\in[-a,a]}|\psi^{(r)}(\tau)|.
\eeq
But since
\beq
\mu^r \sum_{i=1}^\infty (1-\mu)^{r(i-1)}=\frac{\mu^r}{1-(1-\mu)^r},
\eeq
we conclude that $\phi^{(r)}_{k,\mu}(t;n)$ is majorized by an absolutely convergent series. In view of Weierstrass theorem about uniform convergence~\cite[Theorem 7.10, p.~148]{RudinBook}, this result implies that $\phi^{(r)}_{k,\mu}(t;n)$ converges uniformly as $n\rightarrow\infty$ in any compact interval $[-a,a]$. 
Since this holds for all $r=1,2,\dots$, and since we already showed that $\phi_{k,\mu}(t;n)\rightarrow \phi_{k,\mu}(t)$ in~(\ref{eq:Theorem1_1}), the results about uniform convergence and differentiability~\cite[Theorem 7.17, p.~152]{RudinBook} allow us to conclude that the derivative of the limit function equals the limit of the series of derivatives, namely, that~(\ref{eq:Theorem1_3}) holds true.
Moreover, using~(\ref{eq:phiderivativestn}) for $n\rightarrow\infty$, we can write for any $t\neq 0$:
\beq
\phi^{(r)}_{k,\mu}(t)=\frac{1}{t^r}\sum_{i=1}^\infty\sum_{\ell=1}^S (\xi_{i,\ell} t)^r \psi^{(r)}(\xi_{i,\ell}t).
\eeq
Applying Lemma 1 to the function $f(t)=t^r \psi^{(r)}(t)$, with the same choices done before for $b_i$ and $p$, Eq.~(\ref{eq:mainlemmaclaim}) gives:
\beq
\phi^{(r)}_{k,\mu}(t)=\frac{1}{\mu t^r}
\sum_{\ell=1}^S \int_{0}^{p_\ell\mu t} \psi^{(r)}(\tau) \tau^{r-1} d\tau + \frac{1}{t^r}
\sum_{\ell=1}^S R_\ell(t),
\eeq 
which implies:
\beq
\frac{\phi^{(r)}_{k,\mu}(t/\mu)}{\mu^{r-1}}=
\frac{1}{t^r}
\sum_{\ell=1}^S \int_{0}^{p_\ell t}\psi^{(r)}(\tau)\tau^{r-1}d\tau + \frac{\mu}{t^r}\sum_{\ell=1}^S R_\ell(t/\mu),
\label{eq:phikmurthderivo}
\eeq 
and, using~(\ref{eq:Restimate}), the remainder term vanishes as $\mu\rightarrow 0$, implying that
\beq
\frac{\phi_{k,\mu}^{(r)}(t/\mu)}{\mu^{r-1}}\stackrel{\mu\rightarrow 0}{\longrightarrow}
\frac{1}{t^r}\sum_{\ell=1}^S \int_{0}^{p_\ell t} \psi^{(r)}(\tau)\tau^{r-1}d\tau.
\label{eq:currequat1}
\eeq
Now, in view of Lemma 2, and by the definition of $\phi(t)$ in~(\ref{eq:Theorem1_2}), we have:
\beq
\phi^{(r)}(t)=\sum_{\ell=1}^S p_{\ell}^r \, \omega^{(r)}(p_\ell t)=
\frac{1}{t^r}\sum_{\ell=1}^S \int_{0}^{p_\ell t} \psi^{(r)}(\tau)\tau^{r-1}d\tau.
\label{eq:currequat2}
\eeq
Comparing~(\ref{eq:currequat1}) and~(\ref{eq:currequat2}) we arrive at~(\ref{eq:Theorem1_4}) for all $t\neq 0$. 
It remains to consider the case $t=0$, namely, to show that: 
\beq
\frac{\phi_{k,\mu}^{(r)}(0)}{\mu^{r-1}}\stackrel{\mu\rightarrow 0}{\longrightarrow} \phi^{(r)}(0)=
\frac{\psi^{(r)}(0)}{r}\sum_{\ell=1}^S p_\ell^r,
\label{eq:caset0}
\eeq 
where the last equality follows from~(\ref{eq:omega0derivo}).
This equality is established as follows:
\beqa
\phi_{k,\mu}^{(r)}(0)&\stackrel{(\ref{eq:Theorem1_3})}{=}&\psi^{(r)}(0)\sum_{i=1}^\infty\sum_{\ell=1}^S \xi_{i,\ell}^r\nonumber\\
&\stackrel{(\ref{eq:csidef})}{=}&\psi^{(r)}(0)\mu^r \sum_{i=1}^\infty\sum_{\ell=1}^S (1-\mu)^{r(i-1)} b^r_{k,\ell}(i)\nonumber\\
&=&\psi^{(r)}(0)\mu^r \sum_{i=1}^\infty\sum_{\ell=1}^S p_\ell^r (1-\mu)^{r(i-1)}+\nonumber\\
&&
\underbrace{\psi^{(r)}(0)\mu^r \sum_{i=1}^\infty\sum_{\ell=1}^S (1-\mu)^{r(i-1)} (b_{k,\ell}^r(i)-p_\ell^r)}_{\dfz R}
\nonumber\\
&=&
\psi^{(r)}(0)\frac{\mu^r}{1-(1-\mu)^r}\sum_{\ell=1}^S p_\ell^r + R,
\eeqa
or
\beq
\frac{\phi_{k,\mu}^{(r)}(0)}{\mu^{r-1}}= 
\psi^{(r)}(0)\frac{\mu}{1-(1-\mu)^r}\sum_{\ell=1}^S p_\ell^r + 
\frac{R}{\mu^{r-1}}.
\label{eq:finalfinequa}
\eeq
Using the inequality\footnote{The inequality follows from the factorization, holding for $r\in\mathbb{N}$:
\[
a^r-b^r=(a-b)\sum_{k=0}^{r-1} a^k b^{r-1-k},
\]
and from the fact that $b_{k,\ell}(i)$ and $p_\ell$ are not greater than one.
}
\beq
|b_{k,\ell}^r(i)-p^r_\ell|\leq r |b_{k,\ell}(i)-p_\ell|,
\eeq
and~(\ref{eq:Perron}) we can write
\beq
\frac{|R|}{\mu^{r-1}}\leq
r S {\cal C} \lambda \psi^{(r)}(0)\frac{\mu}{1-[\lambda (1-\mu)^r]}.
\label{eq:RRRbound}
\eeq
Therefore, the second term on the RHS of~(\ref{eq:finalfinequa}) vanishes as $\mu\rightarrow 0$, and by application of L'Hospital's rule~\cite{RudinBook} to the first term we get~(\ref{eq:caset0}).
Finally, examining~(\ref{eq:phikmurthderivo}) and~(\ref{eq:RRRbound}) shows that the ${\cal O}(\mu)$ error estimate in~(\ref{eq:Theorem1_6}) holds true.

\section{}

The proof relies on some properties of moment generating functions, which are collected in Sec.~\ref{sec:MGFprop}, and on a technical lemma, whose proof is deferred to Sec.~\ref{sec:Lemma3}.

\vspace*{5pt}
\noindent
{\em Proof of Theorem 3}. 
By definition, $\theta_{\gamma}$ solves the stationary equation
\beq
\phi^{\prime}(\theta_{\gamma})=\gamma.
\label{eq:stateqsolAppendix}
\eeq
Since $\phi^{\prime\prime}(t)>0$, it follows that $\phi^\prime(t)$ is strictly increasing, and, hence, $\theta_{\gamma}$ is unique. Moreover, from~(\ref{eq:Theorem1_5}) used with $r=1$ we obtain:
\beq
\phi^{\prime}(0)=\psi^{\prime}(0)=\E[\bx],
\eeq
recalling that the first derivative of the LMGF, evaluated at the origin, gives the expectation of the random variable~\cite{Dembo-Zeitouni,DenHollander}. The equality $\phi^{\prime}(0)=\E[\bx]$ implies that $\theta_{\gamma}>0$, because by assumption $\gamma>\E[\bx]$, and $\phi^\prime(t)$ is strictly increasing.

For ease of notation, in the following we shall suppress the subscript $\gamma$, and we shall use the symbol $\theta$ to denote the unique solution of~(\ref{eq:stateqsolAppendix}).
Let us denote by $m(dy)$ the probability measure associated with $\by^\star_{k,\mu}$, and introduce the following measure transformation, namely, an exponential translation in the form: 
\beq
\tilde m(dy)=e^{\frac{\theta}{\mu}\,y-\phi_{k,\mu}\left(\frac\theta\mu\right)} m(dy).
\label{eq:exptraappB}
\eeq
The use of a similar transformation for large deviations analysis was originally proposed by Cram\'er~\cite{CramerFundamental}, and then used, e.g., in~\cite{BahadurRao}, to find the exact asymptotics in the simplest case of normalized sums of i.i.d. random variables --- see also~\cite{Dembo-Zeitouni,DenHollander}. We will explain how this transformation is also  helpful in our more general case, where the asymptotic parameter $\mu$ is not simply the index of a summation of i.i.d. random variables.

Denoting by ${\cal I}_{{\cal E}}$ the indicator of an event ${\cal E}$, and by $\E_{\tilde m}[\cdot]$ the expectation operator under the measure $\tilde m(dy)$, some straightforward algebra allows us to write:
\beqa
\lefteqn{
\P[\by^\star_{k,\mu}> \gamma]=\E[{\cal I}_{\{\by^\star_{k,\mu}>\gamma\}}]
=\int_{\gamma}^{\infty}m(dy)}\nonumber\\
&=&
\int_{\gamma}^{\infty}e^{-\frac{\theta}{\mu}\,y+\phi_{k,\mu}\left(\frac\theta\mu\right)}
\tilde m(dy)
\nonumber\\
&=&\E_{\tilde m}[e^{-\frac{\theta}{\mu}\,\by^\star_{k,\mu}+\phi_{k,\mu}\left(\frac\theta\mu\right)}\,{\cal I}_{\{\by^\star_{k,\mu}>\gamma\}}]\nonumber\\
&=&
e^{-\frac 1 \mu [\gamma\theta-\phi(\theta)]}
\,
e^{-\frac 1 \mu \left[\phi(\theta)-\mu\phi_{k,\mu}\left(\frac \theta \mu \right) \right]}\,
e^{\frac\theta\mu\left[\gamma-\phi^\prime_{k,\mu}\left(\frac\theta\mu\right)\right]}\times\nonumber\\
&&\E_{\tilde m}[e^{-\frac{\theta}{\mu}\,[\by^\star_{k,\mu}-\phi^\prime_{k,\mu}\left(\frac\theta\mu\right)]}\,{\cal I}_{\{\by^\star_{k,\mu}>\gamma\}}]\nonumber\\
&=&
e^{-\frac 1 \mu [\Phi(\gamma)+\epsilon_{k,\mu}(\theta)]}
\,
e^{\frac\theta\mu\left[\gamma-\phi^\prime_{k,\mu}\left(\frac\theta\mu\right)\right]}
\nonumber\times\\
&&\E_{\tilde m}[e^{-\frac{\theta}{\mu}\,[\by^\star_{k,\mu}-\phi^\prime_{k,\mu}\left(\frac\theta\mu\right)]}\,{\cal I}_{\{\by^\star_{k,\mu}>\gamma\}}],
\label{eq:chain1}
\eeqa
where we used definition~(\ref{eq:epsilonkmumainexpression}) for $\epsilon_{k,\mu}(\theta)$, and the fact that
\beq
\Phi(\gamma)=\sup_{t\in\mathbb{R}}[\gamma t -\phi(t)]
\Leftrightarrow
\Phi(\gamma)=\gamma \theta - \phi(\theta),
\label{eq:Phigammaderivat}
\eeq 
which holds true because $\phi(t)$ is strictly convex and $\theta$ solves the stationary equation~(\ref{eq:stateqsolAppendix}). It is now useful to introduce the normalized random variable
\beq
\bw^\star_{k,\mu}\dfz\frac{\by^\star_{k,\mu}-\phi^\prime_{k,\mu}(\theta/\mu)}{\sqrt{\phi^{\prime\prime}_{k,\mu}(\theta/\mu)}}.
\label{eq:wstarfirstdef}
\eeq
For ease of notation, we introduce the quantity
\beq
s_{k,\mu}\dfz\sqrt{\frac{\phi_{k,\mu}^{\prime\prime}(\theta/\mu)}{\mu}}\stackrel{\mu\rightarrow 0}{\longrightarrow} \sqrt{\phi^{\prime\prime}(\theta)},
\label{eq:skmuconv}
\eeq
where the convergence follows from~(\ref{eq:Theorem1_4}). With this definition, the expression for $\bw^\star_{k,\mu}$ can be rewritten as:
\beq
\bw^\star_{k,\mu}=\frac{\by^\star_{k,\mu}-\phi^\prime_{k,\mu}(\theta/\mu)}{\sqrt{\mu}s_{k,\mu}}.
\label{eq:newwstar}
\eeq
It is also convenient to introduce the normalized threshold:
\beq
\gamma_{k,\mu}\dfz\frac{\gamma-\phi^\prime_{k,\mu}(\theta/\mu)}{\sqrt{\mu}s_{k,\mu}},
\label{eq:normalizedthresholddef}
\eeq
and the normalized expectation:
\beq
C_{k,\mu}\dfz
\frac{1}{\sqrt{\mu}}\,\E_{\tilde m}\left[
e^{-\frac{\theta s_{k,\mu}}{\sqrt{\mu}}\,(\bw^\star_{k,\mu}-\gamma_{k,\mu})}\,{\cal I}_{\{\bw^\star_{k,\mu}>\gamma_{k,\mu}\}}
\right].
\label{eq:Ckmudefiniz}
\eeq
Using the above definitions, Eq.~(\ref{eq:chain1}) can be rewritten as:
\beq
\P[\by^\star_{k,\mu}>\gamma]=
\sqrt{\mu}\,C_{k,\mu}\,
e^{-\frac 1 \mu [\Phi(\gamma)+\epsilon_{k,\mu}(\theta)]}.
\label{eq:mainexactasywithCkmu}
\eeq
Examining expression~(\ref{eq:mainPkmu_theorem}) for $\mathscr{P}_{k,\mu}(\gamma)$, it is immediate to recognize that the theorem will be established if we are able to show that:
\beq
\boxed{
C_{k,\mu}\stackrel{\mu\rightarrow 0}{\longrightarrow}\frac{1}{\sqrt{2\pi\theta^2 \phi^{\prime\prime}(\theta)}}
}
\label{eq:fundamentalmente}
\eeq
The key for proving the above claim is the asymptotic normality of the random variable $\bw^\star_{k,\mu}$ under the transformed measure $\tilde m(dy)$. To see this, let us exploit the fundamental properties of $\bw^\star_{k,\mu}$. First, according to property P2 in Sec.~\ref{sec:MGFprop}, and since the first three central moments of a random variable coincide with its first three cumulants~\cite{BhattacharyaRao}, under the considered transformed measure we have:
\beqa
\E_{\tilde m}[\by^\star_{k,\mu}]&=&\phi^{\prime}_{k,\mu}(\theta/\mu),\\
\E_{\tilde m}[(\by^\star_{k,\mu}-\phi^{\prime}_{k,\mu}(\theta/\mu))^2]&=&\phi^{\prime\prime}_{k,\mu}(\theta/\mu),\\
\E_{\tilde m}[(\by^\star_{k,\mu}-\phi^{\prime}_{k,\mu}(\theta/\mu))^3]&=&\phi^{\prime\prime\prime}_{k,\mu}(\theta/\mu).
\eeqa
Accordingly, from~(\ref{eq:wstarfirstdef}) and~(\ref{eq:skmuconv}), we conclude that $\bw^\star_{k,\mu}$ is zero-mean, unit-variance, and its third moment is:
\beq
\E_{\tilde m}[(\bw^\star_{k,\mu})^3]=\frac{\phi^{\prime\prime\prime}_{k,\mu}(\theta/\mu)}{\mu^{3/2} s^3_{k,\mu}}.
\eeq
A crucial step consists now in finding an asymptotic expansion for the cumulative distribution function of the random variable $\bw^\star_{k,\mu}$ under the transformed measure. 
To this aim, we follow the procedure employed for deriving the Berry-Esseen and Edgeworth expansions --- see, e.g.,~\cite{FellerBookV2}, and we introduce the function: 
\beqa
G_{k,\mu}(w)&\dfz&
N(w)-\frac{\E_{\tilde m}[(\bw^\star_{k,\mu})^3]}{6}(w^2-1)\,n(w)\nonumber\\
&=&
N(w)-\frac{\phi^{\prime\prime\prime}_{k,\mu}(\theta/\mu)}{6\,\mu^{3/2} s_{k,\mu}^3}(w^2-1)\,n(w),
\label{eq:Gkmu}
\eeqa
where $N(w)$ and $n(w)=N^\prime(w)$ are, respectively, the cumulative distribution function and the probability density function of a standard normal. Denoting by $F_{k,\mu}(w)$ the cumulative distribution function of the random variable $\bw^\star_{k,\mu}$ under the transformed measure, in~\ref{sec:Lemma3} we establish that (Lemma 3):
\beq
\frac{1}{\sqrt{\mu}}\sup_{w\in\mathbb{R}}|F_{k,\mu}(w)-G_{k,\mu}(w)|\stackrel{\mu\rightarrow 0}{\longrightarrow} 0.
\label{eq:refinedconv}
\eeq
This implies that $\bw^\star_{k,\mu}$ is asymptotically normal under the transformed measure $\tilde m(dy)$. More importantly, Eq.~(\ref{eq:refinedconv}) provides a refined, uniform estimate of the convergence error, a property that we shall exploit soon to prove~(\ref{eq:fundamentalmente}). To this aim, we evaluate explicitly the expectation in~(\ref{eq:Ckmudefiniz}). Using integration by parts, we obtain:
\beqa
\lefteqn{C_{k,\mu}
=
\frac{1}{\sqrt{\mu}}\int_{\gamma_{k,\mu}}^\infty e^{-\frac{\theta s_{k,\mu}}{\sqrt{\mu}}(w-\gamma_{k,\mu})}  dF_{k,\mu}(w)
}\nonumber\\
&=&
\frac{1}{\sqrt{\mu}}\left[e^{-\frac{\theta s_{k,\mu}}{\sqrt{\mu}}(w-\gamma_{k,\mu})}\,[F_{k,\mu}(w)-F_{k,\mu}(\gamma_{k,\mu})]\right]_{\gamma_{k,\mu}}^\infty+\nonumber\\
&&
\frac{\theta s_{k,\mu}}{\mu}
\int_{\gamma_{k,\mu}}^\infty
e^{-\frac{\theta s_{k,\mu}}{\sqrt{\mu}}(w-\gamma_{k,\mu})}[F_{k,\mu}(w)-F_{k,\mu}(\gamma_{k,\mu})]dw\nonumber\\
&=&
\frac{1}{\sqrt{\mu}}\int_{0}^\infty
e^{-x}\left[F_{k,\mu}(p_{k,\mu}(x))-F_{k,\mu}(\gamma_{k,\mu})\right]dx,
\label{eq:Ckmudef}
\eeqa
where the last equality follows from the change of variable 
\beq
\frac{\theta s_{k,\mu}}{\sqrt{\mu}}(w-\gamma_{k,\mu}) \rightarrow  x,
\eeq
and from the definition:
\beq
p_{k,\mu}(x)\dfz\frac{\sqrt{\mu}}{\theta s_{k,\mu}} x + \gamma_{k,\mu}.
\label{eq:pkmupoint}
\eeq
Let us also introduce the analogue of~(\ref{eq:Ckmudef}) for the normal approximation $G_{k,\mu}(w)$ in~(\ref{eq:Gkmu}), namely,
\beq
D_{k,\mu}\dfz
\frac{1}{\sqrt{\mu}}\int_{0}^\infty
e^{-x}\left[G_{k,\mu}(p_{k,\mu}(x))-G_{k,\mu}(\gamma_{k,\mu})\right]dx.
\label{eq:Dkmudef0}
\eeq
By using Lemma 3, we can write:
\beqa
|C_{k,\mu}-D_{k,\mu}|
&\leq&
\frac{2}{\sqrt{\mu}}\sup_{w\in\mathbb{R}}|F_{k,\mu}(w)-G_{k,\mu}(w)|
\int_{0}^\infty
e^{-x}dx \nonumber\\
&=&\frac{2}{\sqrt{\mu}}\sup_{w\in\mathbb{R}}|F_{k,\mu}(w)-G_{k,\mu}(w)|
\stackrel{\mu\rightarrow 0}{\longrightarrow} 0.\nonumber\\
\eeqa
By virtue of this result, and in order to prove~(\ref{eq:fundamentalmente}), it is enough to show that
\beq
D_{k,\mu}\stackrel{\mu\rightarrow 0}{\longrightarrow}\frac{1}{\sqrt{2\pi\theta^2 \phi^{\prime\prime}(\theta)}}.
\eeq
Substituting definition~(\ref{eq:Gkmu}) for $G_{k,\mu}(w)$ into~(\ref{eq:Dkmudef0}), we have:
\beqa
D_{k,\mu}&=&
\frac{1}{\sqrt{\mu}}\int_{0}^\infty
e^{-x}\left[N(p_{k,\mu}(x))-N(\gamma_{k,\mu})\right]dx+\nonumber\\
&&
\frac{\phi^{\prime\prime\prime}_{k,\mu}(\theta/\mu)}{6\,\mu^2 s^3_{k,\mu}}
\int_{0}^\infty
e^{-x}\times
\nonumber\\
&&
[
n(\gamma_{k,\mu})(\gamma^2_{k,\mu}-1)
-
n(p_{k,\mu}(x))
(p_{k,\mu}^2(x) - 1)
]dx.
\nonumber\\
\label{eq:Dkmu}
\eeqa
Consider first the second term on the RHS.  By~(\ref{eq:Theorem1_6}) we can write (recall from~(\ref{eq:stateqsolAppendix}) that $\phi^\prime(\theta)=\gamma$):
\beq
\phi^{\prime}_{k,\mu}(\theta/\mu)=\phi^{\prime}(\theta)+{\cal O}(\mu)=\gamma+{\cal O}(\mu),
\eeq
which, in view of~(\ref{eq:normalizedthresholddef}),  implies that 
\beq
\lim_{\mu\rightarrow 0} \gamma_{k,\mu}=0.
\label{eq:gammakmuvanishes}
\eeq
We therefore conclude that $\lim_{\mu\rightarrow 0} p_{k,\mu}(x)=0$ from~(\ref{eq:pkmupoint}) and~(\ref{eq:skmuconv}). 
Accordingly, we have:
\beq
\lim_{\mu\rightarrow 0}[n(\gamma_{k,\mu})(\gamma^2_{k,\mu}-1)
-
n(p_{k,\mu}(x))
(p_{k,\mu}^2(x) - 1)
]=0,
\eeq
and, hence, by dominated convergence~\cite{RudinBookComplex} (note that the function $n(w)(w^2-1)$ is bounded), the second term on the RHS in~(\ref{eq:Dkmu}) vanishes as $\mu\rightarrow 0$, having further used the fact that the ratio $\frac{\phi^{\prime\prime\prime}_{k,\mu}(\theta/\mu)}{6\,\mu^2 s^3_{k,\mu}}$ converges in view of~(\ref{eq:Theorem1_4}) and~(\ref{eq:skmuconv}). 

With regards to the first term on the RHS of~(\ref{eq:Dkmu}), by using a second-order Taylor expansion of $N(w)$ around the point $\gamma_{k,\mu}$ we obtain, for an intermediate point $q_{k,\mu}(x)$:
\beqa
N(p_{k,\mu}(x))
&=&N(\gamma_{k,\mu})+n(\gamma_{k,\mu})\left(p_{k,\mu}(x)-\gamma_{k,\mu}\right)+\nonumber\\
&&
n^\prime(q_{k,\mu}(x))
\frac{(p_{k,\mu}(x)-\gamma_{k,\mu})^2}{2}\nonumber\\
&=&
N(\gamma_{k,\mu})+n(\gamma_{k,\mu})\frac{\sqrt{\mu}}{\theta s_{k,\mu}}x+\nonumber\\
&&
n^\prime(q_{k,\mu}(x))
\frac{\mu x^2}{2 (\theta s_{k,\mu})^2},
\eeqa
where we used~(\ref{eq:pkmupoint}). The above expansion then gives:
\beqa
\lefteqn{
\frac{1}{\sqrt{\mu}}\int_{0}^\infty
e^{-x}\left[
N(p_{k,\mu}(x))-N(\gamma_{k,\mu})
\right]dx
}
\nonumber\\
&=&
\frac{n(\gamma_{k,\mu}) }{\theta s_{k,\mu}}\int_{0}^\infty
x e^{-x} dx+\nonumber\\
&&
\frac{\sqrt{\mu}}{2 (\theta s_{k,\mu})^2}\int_{0}^\infty
x^2 e^{-x}n^\prime(q_{k,\mu}(x)) dx.
\label{eq:finaltermine}
\eeqa
The second term on the RHS of~(\ref{eq:finaltermine}) vanishes because
\beqa
\left|
\int_0^{\infty} 
x^2 e^{-x}n^\prime(q_{k,\mu}(x)) dx
\right|
&\leq&
\sup_{w\in\mathbb{R}} |n^\prime(w)|
\int_0^{\infty} x^2 e^{-x}dx\nonumber\\
&=&
2\sup_{w\in\mathbb{R}} |n^\prime(w)|
<\infty.
\eeqa
On the other hand, the first term on the RHS of~(\ref{eq:finaltermine}) satisfies:
\beq
\frac{n(\gamma_{k,\mu})}{\theta s_{k,\mu}}\int_{0}^\infty x e^{-x} dx
=
\frac{n(\gamma_{k,\mu})}{\theta s_{k,\mu}}
\stackrel{\mu\rightarrow 0}{\longrightarrow}
\frac{1}{\sqrt{2\pi \theta^2 \phi^{\prime\prime}(\theta)}},
\label{eq:lastequat}
\eeq
since $\gamma_{k,\mu}$ vanishes by~(\ref{eq:gammakmuvanishes}), $n(0)=1/\sqrt{2\pi}$, $\int_{0}^\infty x e^{-x} dx=1$, and $s_{k,\mu}$ converges to $\sqrt{\phi^{\prime\prime}(\theta)}$ by~(\ref{eq:skmuconv}). The proof of~(\ref{eq:epsilonkmumainexpression}) is now complete.

It remains to justify the alternative expression~(\ref{eq:epsilonkmumainexpression_bis}). The rationale behind our choice is as follows. Examining~(\ref{eq:lastequat}), it is seen that the term $n(\gamma_{k,\mu})$ is replaced by its limiting constant value of $1/\sqrt{2\pi}$. In order to get a better approximation for finite values of $1/\mu$, we can decide to retain the dependence on $\mu$ by including $n(\gamma_{k,\mu})$ in the main formula. 
This amounts to replacing $C_{k,\mu}$ in~(\ref{eq:mainexactasywithCkmu}), not with the limiting value in~(\ref{eq:fundamentalmente}),  but rather with
\beq
\sqrt{2\pi}\,n(\gamma_{k,\mu})\,\frac{1}{\sqrt{2\pi \theta^2 \phi^{\prime\prime}(\theta)}}.
\eeq
By making explicit the definition of $\gamma_{k,\mu}$ in~(\ref{eq:normalizedthresholddef}), and of the Gaussian pdf, the additional term is:
\beq
\sqrt{2\pi}\,n(\gamma_{k,\mu})=
e^{-\frac{[\gamma-\phi^\prime_{k,\mu}(\theta/\mu)]^2}{2 \mu s^2_{k,\mu}}}\sim
e^{-\frac{1}{\mu}\, \frac{[\phi^\prime(\theta)-\phi^\prime_{k,\mu}(\theta/\mu)]^2}{2 \phi^{\prime\prime}(\theta)}},
\label{eq:lastoflast}
\eeq
where, in the last step, we have used $\gamma=\phi^\prime(\theta)$, and, to be consistent with the approximations adopted so far, we have replaced $s^2_{k,\mu}$ with its limiting value $\phi^{\prime\prime}(\theta)$. Examining the correction at the exponent of~(\ref{eq:lastoflast}) we recognize exactly the second term in~(\ref{eq:epsilonkmumainexpression_bis}).

~\hfill$\square$

\subsection{Some general properties of moment generating functions}
\label{sec:MGFprop}
Consider a random variable $\bx$ with Moment Generating Function (MGF), Logarithmic Moment Generating Function (LMGF), and  Characteristic Function (CF) defined respectively, as:
\beq
\nu(t)\dfz \E[e^{t\bx}],\,\,
\psi(t)\dfz \ln \nu(t),\,\,
\varphi(u)\dfz\E[e^{j u\bx}].
\eeq
We stress that, in this specific subsection, $\bx$ is intended to be a {\em generic} random variable, and there is no need here to give it any particular meaning in terms of a local or a global network statistic. We assume that $\nu(t)$ exists for all $t\in\mathbb{R}$. Under this assumption, we can also define the MGF relative to a complex argument $z$, namely:
\beq
\nu(z)\dfz \E[e^{z\bx}],\,\,z\in\mathbb{C},
\eeq
which is the analytic continuation of $\nu(t)$ over the complex plane~\cite{RudinBookComplex}. Note that, for $z=t+j u$:
\beq
|\nu(z)|=|\E[e^{z\bx}]|\leq
\E[|e^{z\bx}|]=\nu(t).
\label{eq:complexrealbound}
\eeq
In this section, we focus on the following measure transformation (exponential measure translation):
\beq
\tilde m(dx)=e^{\eta x -\psi(\eta)} m(dx)=\frac{e^{\eta x}}{\nu(\eta)} m(dx),
\label{eq:exptraappB2}
\eeq
and establish several important properties characterizing the statistical behavior of $\bx$ under the transformed measure $\tilde m(dx)$.
\begin{itemize}
\item
{\em Property P1.} 
The MGF under the transformation is:
\beq
\boxed{
\tilde\nu(t)=\frac{\nu(\eta+t)}{\nu(\eta)}
}
\eeq
Indeed, by definition:
\beq
\tilde \nu(t)\dfz\E_{\tilde m}[e^{t\bx}]=\frac{\E[e^{(\eta+t)\bx}]}{\nu(\eta)}=\frac{\nu(\eta+t)}{\nu(\eta)}.
\label{eq:complexMGF}
\eeq
\item
{\em Property P2.} 
The $r$-th cumulant $\tilde\chi^{(r)}$ of the variable $\bx$ under the transformation is
\beq
\boxed{
\tilde\chi^{(r)}=\psi^{(r)}(\eta)
}
\label{eq:cumulexplic}
\eeq
Again, from property P1, the LMGF of $\bx$ under the transformed measure is:
\beq
\tilde\psi(t)\dfz\ln\tilde\nu(t)=\ln \nu(\eta+t) -\ln\nu(\eta)=\psi(\eta+t)-\psi(\eta).
\eeq
Accordingly,  the $r$-th cumulant of the variable $\bx$ under the transformation is
\beq
\tilde\chi^{(r)}=\left.\frac{d^r}{dt^r}\tilde\psi(t)\right|_{t=0}=\psi^{(r)}(\eta).
\eeq
\item
{\em Property P3.}
The CF of the random variable $\bx$ under the transformation is
\beq
\boxed{
\tilde \varphi(u)=\frac{\nu(\eta+j u)}{\nu(\eta)}
}
\eeq
where $\nu(z)$, for a complex $z$, was defined by~(\ref{eq:complexMGF}). 
This result follows from an argument similar to the proof of Property P1.
\item
{\em Property P4}. Consider the shifted random variable 
\beq
\hat \bx=\bx-\psi^\prime(\eta).
\eeq
Since the CF of the shifted-and-scaled random variable $a\bx+b$ is 
\beq
\E[e^{j t (a\bx+b)}]=\varphi(a t)e^{j bt}, 
\label{eq:shiftscaleCF}
\eeq
the CF of the shifted variable $\hat \bx$ under the transformed measure is readily computed as:
\beq
\hat\varphi(u)=\tilde\varphi(u) e^{-j\psi^\prime(\eta)u}=\frac{\nu(\eta+j u)}{\nu(\eta)}e^{-j\psi^\prime(\eta)u}.
\eeq
Assume that the parameter $\eta$ ruling the measure transformation lies in a bounded interval, say, $|\eta|\leq\eta_{\max}$. Then, for sufficiently small $\delta>0$, a positive constant $M=M(\delta,\eta_{\max})$ exists such that for $|u|\leq \delta$, and for any choice of $|\eta|\leq\eta_{\max}$, the following expansion holds for the CF of the shifted random variable $\hat\bx$:
\beq
\boxed{
\log\hat\varphi(u)=-\psi^{\prime\prime}(\eta)\frac{u^2}{2}+
\psi^{\prime\prime\prime}(\eta)\frac{(j u)^3}{6}+R(u)
}
\label{eq:Property4mainclaim}
\eeq
with the remainder bounded as:
\beq
|R(u)|\leq M\frac{u^4}{24}.
\label{eq:P4remaind}
\eeq

\vspace*{10pt}
In order to prove the above property, let us introduce the logarithm of a complex number $z$:
\beq
\log z=\ln|z|+j\arg(z).
\eeq
For $|z|<1$, it can be alternatively defined by the Taylor series~\cite{RudinBookComplex}: 
\beq
\log(1+z)=\sum_{n=1}^\infty \frac{(-1)^{n+1}}{n}\,z^n.
\label{eq:logseriesexp}
\eeq
We shall now consider conditions under which the function $\log\hat\varphi(u)$ can be represented by a Taylor series.

Since $\hat\varphi(u)$ is the CF of a zero-mean random variable, the following known bound is obtained from the Taylor expansion of the complex exponential~\cite[Eq.~(4.14), p.~514]{FellerBookV2}:
\beq
|\hat\varphi(u)-1|\leq \E_{\tilde m}[\hat\bx^2]\frac{u^2}{2}=\psi^{\prime\prime}(\eta)\frac{u^2}{2},
\label{eq:CFbound}
\eeq 
where in the last equality we used property P2. We now select a parameter $\delta>0$ such that
\beq
\delta^2<\frac{2}{\displaystyle{\max_{|\eta|\leq \eta_{\max}}}\psi^{\prime\prime}(\eta)}, 
\label{eq:choiceofdelta}
\eeq
where the above choice is meaningful because $\psi(\eta)$ is a LMGF, which implies that  $\psi^{\prime\prime}(\eta)>0$, and that $\psi^{\prime\prime}(\eta)$ is a continuous function, and thus admits a maximum within the interval $[-\eta_{\max},\eta_{\max}]$.
Therefore, over the range $|u|\leq \delta$, Eqs.~(\ref{eq:CFbound}) and~(\ref{eq:choiceofdelta}) give:
\beq
|\hat\varphi(u)-1|< 1.
\label{eq:desiredcond}
\eeq
This allows representing, in the considered range for $u$, the function $\log\hat\varphi(u)$ via its Taylor series. By the definition of cumulants, this series can be written as~\cite[Eq.~(6.8), p.~45]{BhattacharyaRao}:
\beq
\log\hat\varphi(u)=\sum_{r=2}^n \tilde\chi^{(r)}\frac{(j u)^r}{r!} + o(|u|^n),\quad (u\rightarrow 0).
\eeq
Note that, in the above formula, the series coefficients should be given by the cumulants of $\hat\bx$ (computed under the transformed measure). As it can be seen, we used instead the cumulants $\tilde\chi^{(r)}$ of $\bx$. This is because $\hat\bx$ is a shifted version of $\bx$, and so its cumulants are (but for the first one, which is zero) the same as those of $\bx$. 

Consider now the three-term series. Using the explicit formulas for the cumulants under the transformed measure provided by~(\ref{eq:cumulexplic}), we get:
\beq
\log\hat\varphi(u)=-\psi^{\prime\prime}(\eta)\frac{u^2}{2}+
\psi^{\prime\prime\prime}(\eta)\frac{(j u)^3}{6}
+r(\bar u)\frac{u^4}{24},
\label{eq:3termsexpanlog}
\eeq
where $\bar u\in(-\delta,\delta)$, and the function $r(u)$ can be expressed as:
\beqa
r(u)&=&\frac{d^4}{du^4}\log\hat\varphi(u)\nonumber\\
&=&\frac{d^4}{du^4}\log\left[\frac{\nu(\eta+j u)}{\nu(\eta)}e^{-j\psi^\prime(\eta) u}\right],
\eeqa
which leads to (the argument $\eta+j u$ is omitted for ease of notation):
\beqa
r(u)&=&\frac{1}{\nu^4}\left[-6 (\nu^\prime)^4
+12 \nu(\nu^\prime)^2  \nu^{\prime\prime}
- 4 \nu^2\nu^{\prime}\nu^{\prime\prime\prime}-\right.\nonumber\\
&& \left. 3 \nu^2(\nu^{\prime\prime})^2
+ \nu^3 \nu^{(4)}
\right].
\label{eq:4der}
\eeqa
We now show that $|r(u)|$ is bounded for $|\eta|\leq \eta_{\max}$ and $|u|\leq\delta$. To this aim, it is useful to introduce the region of the complex plane:
\beq
{\cal A}=\{\textnormal{Re}(z)\in[-\eta_{\max},\eta_{\max}], \textnormal{Im}(z)\in[-\delta,\delta]\}.
\eeq
First, note that, since all the derivatives of $\nu(z)$ are analytic (and, hence, continuous) in  $\mathbb{C}$~\cite{RudinBookComplex}, the term between brackets in~(\ref{eq:4der}) is bounded within the bounded and closed set ${\cal A}$. 
In addition, for $z\in{\cal A}$, Eq.~(\ref{eq:desiredcond}) implies that $|\nu(z)|>0$. Indeed, if we had $\nu(z)=0$ for a certain $z\in{\cal A}$, then there would exist $\eta$ and $u$, with $|\eta|\leq\eta_{\max}$ and $|u|\leq\delta$, such that $\nu(\eta+j u)=0$, implying:
\beq
|\hat\varphi(u)-1|=\left|\frac{\nu(\eta+j u)}{\nu(\eta)} e^{-j\psi^\prime(\eta)u} -1\right|=1,
\eeq
and this condition would violate~(\ref{eq:desiredcond}).
We can accordingly write, for all $|\eta|\leq\eta_{\max}$ and $|u|\leq\delta$:
\beq
\left|\frac{1}{\nu(\eta+j u)}\right|\leq \max_{z\in{\cal A}} \left|\frac{1}{\nu(z)}\right|<\infty,
\eeq
where finiteness follows since $\nu(z)$ is analytic (and, hence, continuous) in $\mathbb{C}$, and it does not vanish within the set ${\cal A}$. 
We have in fact shown that $|r(u)|$ is bounded for $|\eta|\leq \eta_{\max}$ and $|u|\leq\delta$, namely, that a constant $M=M(\delta,\eta_{\max})$ exists such that $|r(\bar u)|\leq M$. In the light of~(\ref{eq:3termsexpanlog}), this completes the proof of~(\ref{eq:Property4mainclaim}) and~(\ref{eq:P4remaind}).

\end{itemize}

\subsection{Technical lemma relevant to Theorem 3}
\label{sec:Lemma3}
Let us introduce the main quantities involved in the forthcoming derivation.
\begin{itemize}
\item
The measure transformation:
\beq
\tilde m(dy)=e^{\frac{\theta}{\mu}\,y-\phi_{k,\mu}\left(\frac\theta\mu\right)} m(dy).
\label{eq:meastransf33}
\eeq
\item
The random variable introduced in~(\ref{eq:newwstar}):
\beq
\bw^{\star}_{k,\mu}=\frac{\by^\star_{k,\mu}-\phi^{\prime}_{k,\mu}(\theta/\mu)}{\sqrt{\mu}s_{k,\mu}},
\eeq
whose cumulative distribution function under the transformed measure has been denoted by $F_{k,\mu}(w)$.
Using property P3 and applying~(\ref{eq:shiftscaleCF}), it is easily seen that the CF of $\bw^\star_{k,\mu}$ under the transformed measure is:
\beq
\tilde\varphi_{k,\mu}(u)=
\frac{\nu_{k,\mu}\left(\frac{\theta}{\mu}  + j \frac{u}{\sqrt{\mu}s_{k,\mu}}\right)}
{\nu_{k,\mu}\left(\frac{\theta}{\mu}\right)}\,
e^{-j u\frac{\phi^{\prime}_{k,\mu}(\theta/\mu)}{\sqrt{\mu}s_{k,\mu}}},
\label{eq:tildephi}
\eeq
where $\nu_{k,\mu}(\cdot)$ denotes the MGF (of real or complex argument) of the random variable $\by^\star_{k,\mu}$ under the original measure.
\item
The normal approximation $G_{k,\mu}(w)$ defined by~(\ref{eq:Gkmu}):
\beq
G_{k,\mu}(w)=N(w)-\frac{\phi^{\prime\prime\prime}_{k,\mu}(\theta/\mu)}{6\,\mu^{3/2}s^3_{k,\mu}}(w^2-1)\,n(w).
\label{eq:Gkmu2}
\eeq
\end{itemize}

\noindent
The following lemma shows that $F_{k,\mu}(w)$ converges to $G_{k,\mu}(w)$ as $\mu$ goes to zero. 
From~(\ref{eq:Theorem1_4}) and~(\ref{eq:skmuconv}), we conclude that the second term on the RHS of~(\ref{eq:Gkmu2}) vanishes as $\mu\rightarrow 0$. This result will imply that $\bw^\star_{k,\mu}$ is asymptotically normal. In addition, regarding $G_{k,\mu}(w)$ as a higher-order approximation for $F_{k,\mu}(w)$ (namely, including a correction with respect to the crudest approximation $N(w)$) the lemma provides the useful information that the (worst-case) rate of convergence of the approximation error is in the order of $\sqrt{\mu}$. This observation is critical for the proof of Theorem 3.

\vspace*{5pt}
\noindent
{\bf \textsc{Lemma 3} (Asymptotic normality of $\bw^\star_{k,\mu}$ under the transformed measure~(\ref{eq:meastransf33}): Estimate of the convergence error).} 
{\em 
Under the same hypotheses of Theorem 3:
\beq
\frac{1}{\sqrt{\mu}}\sup_{w\in\mathbb{R}}|F_{k,\mu}(w)-G_{k,\mu}(w)|\stackrel{\mu\rightarrow 0}{\longrightarrow} 0.
\label{eq:Lemma3Mainclaim}
\eeq
}

\vspace*{5pt}
\noindent
{\em Proof}. 
The following classical result~\cite[page 538]{FellerBookV2} is used in our proof.

\vspace*{5pt}
\noindent
{\em Let $F$ be a probability distribution with zero mean and characteristic function $\varphi$. Suppose that $F-G$ vanishes at $\pm\infty$ and that $G$ has a derivative $g$ such that $|g|\leq m$. Finally, suppose that $g$ has a continuously differentiable Fourier transform $\zeta$ such that $\zeta(0)=1$ and $\zeta^\prime(0)=0$. Then, for all $w$ and for all $T>0$:
\beq
|F(w)-G(w)|\leq\frac{1}{\pi}\int_{-T}^T \left|\frac{\varphi(u)-\zeta(u)}{u}\right|du+\frac{24 m}{\pi T}.
\label{eq:FellerLemma2}
\eeq
}

\vspace*{10pt}
\noindent
It is now immediate to verify that the functions $F_{k,\mu}(w)$ and $G_{k,\mu}(w)$ introduced above meet the conditions required for~(\ref{eq:FellerLemma2}) to hold. Indeed, $G_{k,\mu}(w)\rightarrow 0$ as $w\rightarrow -\infty$, and $G_{k,\mu}(w)\rightarrow 1$ as $w\rightarrow +\infty$, implying that the difference $F_{k,\mu}(w)-G_{k,\mu}(w)$ vanishes at $\pm \infty$ because $F_{k,\mu}(w)$ is a cumulative distribution function. Moreover,
\beqa
\frac{d}{dw}G_{k,\mu}(w)&=&
n(w)\left[1-w\,\frac{\phi^{\prime\prime\prime}_{k,\mu}(\theta/\mu)}{3\,\mu^{3/2} s_{k,\mu}^3}\right]-\nonumber\\
&&
\frac{\phi^{\prime\prime\prime}_{k,\mu}(\theta/\mu)}{6\,\mu^{3/2} s_{k,\mu}^3}(w^2-1)\,n^\prime(w).
\label{eq:Gkmuprime}
\eeqa
From~(\ref{eq:Theorem1_4}) used with $r=3$, we conclude that the quantity $\frac{\phi^{\prime\prime\prime}_{k,\mu}(\theta/\mu)}{\mu^{3/2} s_{k,\mu}^3}$ vanishes as $\mu\rightarrow 0$ and, hence, it is bounded for sufficiently small values of $\mu$. The required boundedness condition on $\left|\frac{d}{dw}G_{k,\mu}(w)\right|$ then follows by the boundedness of the functions $n(w)$, $w\,n(w)$, and $(w^2-1)n^\prime(w)$. Finally, the Fourier transform of~(\ref{eq:Gkmuprime}) is given by~\cite{FellerBookV2}:
\beq
\zeta_{k,\mu}(u)\dfz e^{-\frac{u^2}{2}}\left[
1+\frac{\phi^{\prime\prime\prime}_{k,\mu}(\theta/\mu)}{6\,\mu^{3/2}s^3_{k,\mu}}(j u)^3
\right].
\label{eq:zetakmudef}
\eeq
It is seen that $\zeta_{k,\mu}(u)$ is continuously differentiable, and that $\zeta_{k,\mu}(0)=1$, $\zeta^\prime_{k,\mu}(0)=0$.

In order to use the above lemma, let us select, for a given $\epsilon>0$, a constant $a$ such that:
\beq
a>\frac{24 m}{\epsilon},
\eeq
and set
\beq
T=\frac{a}{\sqrt{\mu}}\Rightarrow \frac{24 m}{\pi T}\leq \epsilon\sqrt{\mu}.
\eeq
With these particular choices, application of~(\ref{eq:FellerLemma2}) yields:
\beqa
\lefteqn{\frac{1}{\sqrt{\mu}}|F_{k,\mu}(w)-G_{k,\mu}(w)|}\nonumber\\
&\leq&
\frac{1}{\sqrt{\mu}}
\int_{-\frac{a}{\sqrt{\mu}}}^{\frac{a}{\sqrt{\mu}}} \left|\frac{\tilde\varphi_{k,\mu}(u)-\zeta_{k,\mu}(u)}{u}\right|du+\epsilon.
\label{eq:applicationofFeller}
\eeqa
Due to arbitrariness of $\epsilon$, the claim~(\ref{eq:Lemma3Mainclaim}) will be proved if we show that the first term on the RHS of~(\ref{eq:applicationofFeller}) vanishes as $\mu\rightarrow 0$. To get this result, we split the integral as follows:
\beq
\frac{1}{\sqrt{\mu}}
\int_{-\frac{a}{\sqrt{\mu}}}^{\frac{a}{\sqrt{\mu}}} \left|\frac{\tilde\varphi_{k,\mu}(u)-\zeta_{k,\mu}(u)}{u}\right|du
=I_1+I_2,
\eeq
where
\beq
I_1=
\frac{1}{\sqrt{\mu}}
\int_{-\frac{\delta}{\sqrt{\mu}}}^{\frac{\delta}{\sqrt{\mu}}} \left|\frac{\tilde\varphi_{k,\mu}(u)-\zeta_{k,\mu}(u)}{u}\right|du,
\label{eq:I1}
\eeq
and
\beq
I_2=
\frac{1}{\sqrt{\mu}}
\int_{\frac{\delta}{\sqrt{\mu}}<|u|<\frac{a}{\sqrt{\mu}}} \left|\frac{\tilde\varphi_{k,\mu}(u)-\zeta_{k,\mu}(u)}{u}\right|du,
\label{eq:I2}
\eeq
for a sufficiently small $\delta$ whose choice will be carefully addressed in the following. We start by proving that $I_1$ vanishes as $\mu\rightarrow 0$.

\vspace*{10pt}
\noindent
{\em Part $i$ ($I_1\rightarrow 0$ as $\mu\rightarrow 0$}). 
From~(\ref{eq:Theorem1_1}) we know that the MGF (with real argument $t$) of the random variable $\by^\star_{k,\mu}$ under the original measure can be represented as:
\beq
\nu_{k,\mu}(t)=\prod_{i=1}^\infty
\prod_{\ell=1}^S
\nu\left(\mu (1-\mu)^{i-1}b_{k,\ell}(i) t\right),
\label{eq:infprod}
\eeq
where $\nu(t)=\E[e^{t \bx_k(n)}]$ is the MGF of the local statistic $\bx_k(n)$.
We also know that the MGF $\nu_{k,\mu}(t)$ can be extended to the complex plane by considering
\beq
\nu_{k,\mu}(z)=\E[e^{z \by^\star_{k,\mu}}]
\eeq 
for complex $z$. It would be useful to have the same infinite-product representation as in~(\ref{eq:infprod}) for $\nu_{k,\mu}(z)$, namely:
\beq
\nu_{k,\mu}(z)=\prod_{i=1}^\infty
\prod_{\ell=1}^S
\nu\left(\mu (1-\mu)^{i-1}b_{k,\ell}(i) z\right).
\label{eq:finalcomplex}
\eeq
This turns out to be true, and can be justified by an argument similar to that used in~\cite{CurtissMGF}, which is now illustrated in detail.  Consider the sequence of functions, for $n=1,2,\dots$
\beq
f_n(z)=\prod_{i=1}^n
\prod_{\ell=1}^S
\nu\left(\mu (1-\mu)^{i-1}b_{k,\ell}(i) z\right).
\eeq
First, $f_n(z)$ is analytic in $\mathbb{C}$, since it consists of the product of functions $\nu(\cdot)$ that are analytic in $\mathbb{C}$. 
Moreover, by~(\ref{eq:complexrealbound}), we can write, for $z=t+j u$:
\beq
|f_n(z)|\leq 
\prod_{i=1}^n
\prod_{\ell=1}^S
\nu\left(\mu (1-\mu)^{i-1}b_{k,\ell}(i) t\right)=f_n(t).
\eeq 
For any $n$, the RHS is a MGF and, hence, its second derivative over $t$ is strictly positive, which implies that the maximum of the RHS within an interval $|t|\leq t_{\max}$ is attained at either one or both endpoints of the interval, namely, that:
\beq
|f_n(z)|\leq\max\left\{f_n(-t_{\max}), f_n(t_{\max})\right\}.
\eeq
But since $f_n(t)$ converges to $\nu_{k,\mu}(t)<\infty$ as $n\rightarrow\infty$, we have that, for any $\epsilon>0$, a certain $n_0$ exists such that, for all $n\geq n_0$:
\beq
f_n(-t_{\max})\leq \nu_{k,\mu}(-t_{\max})+\epsilon,
~~
f_n(t_{\max})\leq \nu_{k,\mu}(t_{\max})+\epsilon,
\eeq
which shows that the sequence $|f_n(z)|$ is uniformly bounded, at least for $n\geq n_0$, in the closed strip $\{z\in\mathbb{C}: |\textnormal{Re}(z)|\leq t_{\max}\}$. 
Moreover, since the sequence $f_n(t)$ ($t\in\mathbb{R}$) admits a limit, we can say that $f_n(z)$ ($z\in\mathbb{C}$) admits a limit at each point of the real-axis, namely, of the interval $\{z\in\mathbb{C}: |\textnormal{Re}(z)|\leq t_{\max},\,\textnormal{Im}(z)=0\}$. This allows invoking Vitali's convergence theorem~\cite{TitchsmarshTOF}, which asserts that $f_n(z)$ converges, namely:
\beq
f_n(z)\stackrel{n\rightarrow\infty}{\longrightarrow} f(z)=
\prod_{i=1}^\infty
\prod_{\ell=1}^S
\nu\left(\mu (1-\mu)^{i-1}b_{k,\ell}(i) z\right),
\eeq
uniformly in each bounded closed subregion of the open strip $\{z\in\mathbb{C}: |\textnormal{Re}(z)|< t_{\max}\}$, and the limit function $f(z)$ is therein analytic. Since we are free to choose $t_{\max}$, we conclude that $f(z)$ is analytic in $\mathbb{C}$. Moreover, on the real axis
\beq
f(t)=
\prod_{i=1}^\infty
\prod_{\ell=1}^S
\nu\left(\mu (1-\mu)^{i-1}b_{k,\ell}(i) t\right)
=\nu_{k,\mu}(t),
\eeq
and since $\nu_{k,\mu}(z)$ is analytic and equals $\nu_{k,\mu}(t)$ on the real axis, by~\cite[Corollary, p.~209]{RudinBookComplex} we finally get~(\ref{eq:finalcomplex}).

\vspace*{10pt}
In the light of the above finding, we are allowed to use~(\ref{eq:finalcomplex}) into~(\ref{eq:tildephi}). This yields, after some manipulations:
\beq
\tilde\varphi_{k,\mu}(u)=
\prod_{i=1}^\infty
\prod_{\ell=1}^S
\frac{\nu(\eta_{i,\ell} + j u_{i,\ell})}
{\nu(\eta_{i,\ell})}
e^{-j \psi^\prime(\eta_{i,\ell})u_{i,\ell} },
\label{eq:CFtildebasic}
\eeq
where we introduced
\beqa
\eta_{i,\ell}&\dfz&(1-\mu)^{i-1}b_{k,\ell}(i)\,\theta,
\label{eq:tauielle}
\\
u_{i,\ell}&\dfz&(1-\mu)^{i-1}b_{k,\ell}(i)\frac{\sqrt{\mu} u}{s_{k,\mu}},
\label{eq:tdefini}
\eeqa
and used~(\ref{eq:Theorem1_3}) to compute $\phi^{\prime}_{k,\mu}(\theta/\mu)$.
Now, from property P4 it is seen that the function
\beq
\frac{\nu(\eta_{i,\ell}+j u_{i,\ell})}
{\nu(\eta_{i,\ell})} e^{-j \psi^\prime(\eta_{i,\ell})u_{i,\ell} }
\eeq
can be regarded as the CF, evaluated at the point $u_{i,\ell}$, of the {\em shifted} random variable
$\bx_\ell(i)-\psi^{\prime}(\eta_{i,\ell})$ under the measure transformation
\beq
\tilde m(dx)=\frac{e^{\eta_{i,\ell} x}}{\nu(\eta_{i,\ell})} m(dx).
\eeq 
We shall denote this CF by $\hat\varphi_{\eta_{i,\ell}}(u_{i,\ell})$, having made explicit the dependence upon $\eta_{i,\ell}$, thus getting:
\beq
\tilde\varphi_{k,\mu}(u)=
\prod_{i=1}^\infty
\prod_{\ell=1}^S
\hat\varphi_{\eta_{i,\ell}}(u_{i,\ell}).
\label{eq:phidoubletilde}
\eeq
From~(\ref{eq:tdefini})  we get: 
\beq
|u_{i,\ell}|\leq \frac{\sqrt{\mu} |u|}{s_{k,\mu}}.
\eeq
We also know from~(\ref{eq:skmuconv}) that  $\lim_{\mu\rightarrow 0} s_{k,\mu}=\sqrt{\phi^{\prime\prime}(\theta)}$. By applying the definition of limit, this implies that, for any $\epsilon>0$, and for sufficiently small values of $\mu$:
\beq
|u_{i,\ell}|\leq
\sqrt{\mu} |u|\left(
\frac{1}{\sqrt{\phi^{\prime\prime}(\theta)}}+\epsilon
\right)
\leq
\delta
\left(
\frac{1}{\sqrt{\phi^{\prime\prime}(\theta)}}+\epsilon
\right)\dfz \bar\delta,
\label{eq:tielleboundbardelta}
\eeq
where the last inequality holds because we are focusing on the integral $I_1$ in~(\ref{eq:I1}) and, hence, we consider the range $\sqrt{\mu}|u|\leq\delta$. 

Noting now that $|\eta_{i,\ell}|\leq\theta$ by~(\ref{eq:tauielle}), the choice
\beq
\bar\delta^2<\frac{2}{\displaystyle{\max_{|\eta|\leq\theta}}\,\psi^{\prime\prime}(\eta)}
\eeq
corresponds to~(\ref{eq:choiceofdelta}), and allows invoking property P4, so that it is legitimate to use the Taylor expansion:
\beqa
\log \hat\varphi_{\eta_{i,\ell}}(u_{i,\ell})&=&
-\psi^{\prime\prime}(\eta_{i,\ell})\frac{u_{i,\ell}^2}{2}
+\psi^{\prime\prime\prime}(\eta_{i,\ell})\frac{(j u_{i,\ell})^3}{6}
+\nonumber\\&&R(u_{i,\ell}).
\eeqa
This leads to
\beqa
&&\lefteqn{
\prod_{i=1}^n
\prod_{\ell=1}^S
\hat\varphi_{\eta_{i,\ell}}(u_{i,\ell})
=
\prod_{i=1}^n
\prod_{\ell=1}^S
e^{\log \hat\varphi_{\eta_{i,\ell}}(u_{i,\ell})
}
}
\nonumber\\
&=&
\prod_{i=1}^n
\prod_{\ell=1}^S
e^{-\psi^{\prime\prime}(\eta_{i,\ell})\frac{u_{i,\ell}^2}{2}+
\psi^{\prime\prime\prime}(\eta_{i,\ell})\frac{(j u_{i,\ell})^3}{6}
+R(u_{i,\ell})}\nonumber\\
&=&
e^{
\sum_{i=1}^n
\sum_{\ell=1}^S
-\psi^{\prime\prime}(\eta_{i,\ell})\frac{u_{i,\ell}^2}{2}+
\psi^{\prime\prime\prime}(\eta_{i,\ell})\frac{(j u_{i,\ell})^3}{6}
+R(u_{i,\ell})}.\nonumber\\
\label{eq:finiteprodCF}
\eeqa
We now show that all terms in the exponent are convergent as $n\rightarrow\infty$. 
By using the expressions for $\eta_{i,\ell}$, $u_{i,\ell}$ in~(\ref{eq:tauielle}) and~(\ref{eq:tdefini}), and recalling the definition of $\xi_{i,\ell}$ in~(\ref{eq:csidef}), straightforward manipulations yield:
\beqa
\sum_{i=1}^\infty
\sum_{\ell=1}^S
\psi^{\prime\prime}(\eta_{i,\ell})\frac{u_{i,\ell}^2}{2}
&=&
\frac{1}{\mu s^2_{k,\mu}}
\sum_{i=1}^\infty
\sum_{\ell=1}^S
\xi_{i,\ell}^2\psi^{\prime\prime}(\xi_{i,\ell}\theta/\mu)\frac{u^2}{2}\nonumber\\
&=&
\frac{\phi^{\prime\prime}_{k,\mu}(\theta/\mu)}{\mu s^2_{k,\mu}}\frac{u^2}{2}=\frac{u^2}{2},
\eeqa
having used, in the last two equalities, Eq.~(\ref{eq:Theorem1_3}) and the definition of $s_{k,\mu}$ in~(\ref{eq:skmuconv}).
Similarly,
\beqa
A_1&\dfz&\sum_{i=1}^\infty
\sum_{\ell=1}^S
\psi^{\prime\prime\prime}(\eta_{i,\ell})\frac{( j u_{i,\ell})^3}{6}
\nonumber\\
&=&
\frac{1}{\mu^{3/2} s^3_{k,\mu}}
\sum_{i=1}^\infty
\sum_{\ell=1}^S
\xi_{i,\ell}^3\psi^{\prime\prime\prime}(\xi_{i,\ell}\theta/\mu)\frac{(j u)^3}{6}\nonumber\\
&=&
\frac{\phi^{\prime\prime\prime}_{k,\mu}(\theta/\mu)}{\mu^{3/2} s^3_{k,\mu}}\frac{(j u)^3}{6}.
\eeqa
Finally, by defining
\beq
A_2\dfz
\sum_{i=1}^\infty
\sum_{\ell=1}^S
R(u_{i,\ell}),
\eeq
and using~(\ref{eq:P4remaind}), we have (recall that $b_{k,\ell}(i)\leq 1$):
\beqa
|A_2|&\leq&
\sum_{i=1}^\infty
\sum_{\ell=1}^S
|R(u_{i,\ell})|
\leq
\frac{M}{24}
\sum_{i=1}^\infty
\sum_{\ell=1}^S
|u_{i,\ell}|^4
\nonumber\\
&\leq&
\frac{M}{24 s^4_{k,\mu}}\mu^2
\sum_{i=1}^\infty
\sum_{\ell=1}^S
(1-\mu)^{4(i-1)}
|u|^4\nonumber\\
&=&
\frac{M S}{24 s^4_{k,\mu}}\frac{\mu^2}{1-(1-\mu)^4} |u|^4.
\eeqa
We stress that the constant $M$ appearing in the above formulas bounds uniformly all terms $|R(u_{i,\ell})|$. Indeed, as already observed, in the considered range of analysis we have $|u_{i,\ell}|\leq\bar\delta$ and $|\eta_{i,\ell}|\leq \theta$. Under this assumption, from property P4 we know that the bounding constant $M$ depends only on the pair $(\bar\delta,\theta)$, and, hence, not on the particular pair $(i,\ell)$.

Using~(\ref{eq:phidoubletilde}) and~(\ref{eq:finiteprodCF}), it is now legitimate to rewrite
\beq
\tilde\varphi_{k,\mu}(u)=e^{-\frac{u^2}{2}+A_1+A_2}.
\eeq
Making explicit the definition of $\zeta_{k,\mu}(u)$ appearing in~(\ref{eq:I1}), we have:
\beqa
\lefteqn{
\left|
\tilde\varphi_{k,\mu}(u)-\zeta_{k,\mu}(u)
\right|
}
\nonumber\\
&=&
\left|
\tilde\varphi_{k,\mu}(u)-e^{-\frac{u^2}{2}}\left(
1+\frac{\phi^{\prime\prime\prime}_{k,\mu}(\theta/\mu)}{6\,\mu^{3/2}s_{k,\mu}^3}(j u)^3
\right)
\right|
\nonumber\\
&=&
e^{-\frac{u^2}{2}}
\left|
\tilde\varphi_{k,\mu}(u)e^{\frac{u^2}{2}}
-1-\frac{\phi^{\prime\prime\prime}_{k,\mu}(\theta/\mu)}{\mu^{3/2}s_{k,\mu}^3}\frac{(j u)^3}{6}
\right|\nonumber\\
&=&
e^{-\frac{u^2}{2}}
\left|e^{A_1+A_2}
-1-A_1\right|\nonumber\\
&\leq&
e^{-\frac{u^2}{2}}
\left(|A_2|+\frac{|A_2|^2}{2}
\right)e^{|A_1|+|A_2|},
\label{eq:integrandchaineq}
\eeqa
where, in the last inequality, we resorted to the following result, which is known to hold for any pair of complex numbers $z$ and $w$ --- see, e.g.,~\cite[Eq.~(2.8), p.~534]{FellerBookV2}:
\beq
|e^{z}-1-w|\leq 
\left(|z-w|+\frac 1 2 |w|^2\right)e^{\max(|z|,|w|)}.
\eeq
Consider first the term $A_1$. 
Using~(\ref{eq:skmuconv}) and~(\ref{eq:Theorem1_4}) with $r=3$, we can write:
\beq
\lim_{\mu\rightarrow 0}\left|
\frac{\phi_{k,\mu}^{\prime\prime\prime}(\theta/\mu)}{\mu^2 s^3_{k,\mu}}
\right|=
\left|
\frac{\phi^{\prime\prime\prime}(\theta)}{[\phi^{\prime\prime}(\theta)]^{3/2}}
\right|.
\eeq
Making explicit the definition of limit, for any $\epsilon>0$, and for $\mu$ small enough, it holds that:
\beqa
|A_1|&=&
\frac{1}{6\mu}
\left|\frac{\phi^{\prime\prime\prime}_{k,\mu}(\theta/\mu)}{\mu^2 s^3_{k,\mu}}\right|
|\sqrt{\mu}u|^3\nonumber\\
&\leq&
\frac{1}{6 \mu}
\left(
\left|
\frac{\phi^{\prime\prime\prime}(\theta)}{[\phi^{\prime\prime}(\theta)]^{3/2}}
\right|
+\epsilon
\right)
|
\sqrt{\mu}u
|^3\nonumber\\
&=&
\frac{1}{\mu}
a_1
|
\sqrt{\mu}u
|^3,
\eeqa
where the constant $a_1$ has been implicitly defined.
We switch to the term $A_2$. Applying L'Hospital's rule~\cite{RudinBook}, and using again~(\ref{eq:skmuconv}), we see that:
\beq
\lim_{\mu\rightarrow 0} 
\left(
\frac{\mu}{1-(1-\mu)^4}\,\frac{1}{s_{k,\mu}^4}
\right)=
\frac{1}{4[\phi^{\prime\prime}(\theta)]^2},
\eeq
implying that, for any $\epsilon>0$, and for sufficiently small values of $\mu$:
\beqa
|A_2|&\leq&
\frac{1}{\mu}\frac{M S}{24 s_{k,\mu}^4}
\frac{\mu}{1-(1-\mu)^4}
(\sqrt{\mu}u)^4\nonumber\\
&\leq&
\frac{1}{\mu}\frac{M S}{24}
\left(\frac{1}{4 [\phi^{\prime\prime}(\theta)]^2}+\epsilon\right)
(\sqrt{\mu}u)^4\nonumber\\
&=&
\frac{1}{\mu}a_2\,
(\sqrt{\mu}u)^4,
\eeqa
where the constant $a_2$ has been implicitly defined.
Observe now that, for any $a>0$, any $0\leq x\leq \delta$, and for any integers $m$ and $n$ such that $m<n$:
\beq
\delta^{n-m}\leq 1/a \Rightarrow a x^n\leq x^m.
\eeq
Setting $x=|\sqrt{\mu} u|$, and recalling that in the considered range $|\sqrt{\mu} u|\leq \delta$, the above relationship implies that the parameter $\delta$ can be certainly chosen to satisfy:
\beq
a_1 |\sqrt{\mu} u|^3\leq\frac{|\sqrt{\mu}u|^2}{8},\quad
a_2 (\sqrt{\mu} u)^4\leq\frac{|\sqrt{\mu}u|^2}{8},
\eeq 
yielding (in the considered range):
\beq
|A_1|\leq \frac{u^2}{8},\qquad |A_2|\leq \frac{u^2}{8}.
\label{eq:finalA1A2}
\eeq
Finally, using these inequalities in~(\ref{eq:integrandchaineq}), the integrand in~(\ref{eq:I1}) can be upper bounded as:
\beq
\frac{1}{\sqrt{\mu}}
\left|
\frac{
\tilde\varphi_{k,\mu}(u)-\zeta_{k,\mu}(u)
}{u}
\right|
\leq
e^{-\frac{u^2}{4}}
\left(
\sqrt{\mu}\,a_2 |u|^3 + \mu^{3/2} \frac{a_2^2 |u|^7}{2}
\right),
\eeq
implying that $I_1\rightarrow 0$ as $\mu\rightarrow 0$.

\vspace*{10pt}
\noindent
{\em Part $ii$ ($I_2\rightarrow 0$ as $\mu\rightarrow 0$)}.
The following chain of inequalities holds true:
\beqa
I_2&\leq&\frac{1}{\sqrt{\mu}}
\int_{\frac{\delta}{\sqrt{\mu}}<|u|<\frac{a}{\sqrt{\mu}}}
\left|
\frac{
\tilde\varphi_{k,\mu}(u)}
{u}
\right|
du+
\nonumber\\
&&
\frac{1}{\sqrt{\mu}}
\int_{|u|>\frac{\delta}{\sqrt{\mu}}}
e^{-\frac{u^2}{2}}
\left|
\frac{
1+\frac{\phi^{\prime\prime\prime}_{k,\mu}(\theta/\mu)}{6\,\mu^{3/2}s_{k,\mu}^3}(j u)^3
}{u}
\right|
du\nonumber\\
&\leq&
\frac{1}{\delta}
\int_{\frac{\delta}{\sqrt{\mu}}<|u|<\frac{a}{\sqrt{\mu}}}
|\tilde\varphi_{k,\mu}(u)|
du+\nonumber\\
&&
\frac{1}{\delta}
\int_{|u|>\frac{\delta}{\sqrt{\mu}}}
e^{-\frac{u^2}{2}}du+\nonumber\\
&&
\frac{|\phi^{\prime\prime\prime}_{k,\mu}(\theta/\mu)|}{6\,\mu^2 s^3_{k,\mu}}
\int_{|u|>\frac{\delta}{\sqrt{\mu}}}
e^{-\frac{u^2}{2}}
u^2
du,
\label{eq:inequchainchain}
\eeqa
which follows by repeated application of the triangle inequality for complex numbers $|z+w|\leq |z|+|w|$. Since the integrals 
\beq
\int_{-\infty}^\infty e^{-\frac{u^2}{2}}du,
\quad
\int_{-\infty}^\infty e^{-\frac{u^2}{2}}u^2 du
\eeq
are convergent, and since the quantities $\phi^{\prime\prime\prime}_{k,\mu}(\theta/\mu)/\mu^2$ and $s_{k,\mu}$ converge in view of~(\ref{eq:Theorem1_4}) and~(\ref{eq:skmuconv}), the last two integrals on the RHS of~(\ref{eq:inequchainchain}) vanish as $\mu\rightarrow 0$.

It remains to prove that the first integral on the RHS of~(\ref{eq:inequchainchain}) vanishes as $\mu\rightarrow 0$. 
To this aim, observe that, by~(\ref{eq:CFtildebasic}): 
\beq
|\tilde\varphi_{k,\mu}(u)|
=
\prod_{i=1}^{\infty}
\prod_{\ell=1}^S
\frac{|\nu(\eta_{i,\ell} + j u_{i,\ell})|}
{\nu(\eta_{i,\ell})}.
\label{eq:CFproduct0}
\eeq
Since, from~(\ref{eq:complexrealbound}), we know that 
\beq
\frac{|\nu(\eta_{i,\ell} + j u_{i,\ell})|}
{\nu(\eta_{i,\ell})}\leq1,
\eeq
all the terms of the product in~(\ref{eq:CFproduct0}) are not greater than one, implying that, in particular:
\beq
|\tilde\varphi_{k,\mu}(u)|
\leq
\prod_{i=\lceil 1/\mu\rceil}^{2\lfloor 1/\mu\rfloor}
\prod_{\ell=1}^S
\frac{|\nu(\eta_{i,\ell} + j u_{i,\ell})|}
{\nu(\eta_{i,\ell})},
\label{eq:CFproduct}
\eeq
where $\lceil x\rceil$ (resp., $\lfloor x\rfloor$) denotes the smallest (resp., the largest) integer not smaller (resp., not greater) than $x$. 
Since in the above product we have $1/\mu\leq \lceil 1/\mu \rceil\leq i\leq 2\lfloor 1/\mu \rfloor\leq 2/\mu$, and since, in the range of interest $\delta<|\sqrt{\mu} u|<a$, for $u_{i,\ell}$ defined in~(\ref{eq:tdefini}) we can write:
\beq
\frac{(1-\mu)^{2/\mu-1}b_{k,\ell}(i)}{s_{k,\mu}}
\,\delta<
|u_{i,\ell}|
<
\frac{(1-\mu)^{1/\mu-1}b_{k,\ell}(i)}{s_{k,\mu}} \,a.
\eeq
The following convergences hold: 
\beqa
\lim_{\mu\rightarrow 0} (1-\mu)^{1/\mu}&=&1/e,~~~~~~~\textnormal{[known limit]}
\label{eq:limit11}\\
\lim_{i\rightarrow \infty} b_{k,\ell}(i)&=&p_{\ell},~~~~~~~~~\textnormal{[Eq.~(\ref{eq:bconv})]}\\
\label{eq:limit22}
\lim_{\mu\rightarrow 0} s_{k,\mu}(i)&=&\sqrt{\phi^{\prime\prime}(\theta)}.~~~\textnormal{[Eq.~(\ref{eq:skmuconv})]}
\label{eq:limit33}
\eeqa
Using the explicit definition of limit, the above three relationships imply that, for any $\epsilon>0$, it is possible to choose $\mu$ sufficiently small so as to ensure that, in the considered range $i\geq \lceil 1/\mu \rceil$:
\beq
v_1\dfz\left(\frac{p_\ell/e^2}{\sqrt{\phi^{\prime\prime}(\theta)}}-\epsilon\right) \delta
<|u_{i,\ell}|<
\left(\frac{p_\ell/e}{\sqrt{\phi^{\prime\prime}(\theta)}}+\epsilon\right) a
\dfz v_2,
\eeq
where, due to arbitrariness of $\epsilon$, we can safely assume that $v_1>0$.

Moreover, for $\eta_{i,\ell}$ defined in~(\ref{eq:tauielle}), we have $0\leq\eta_{i,\ell}\leq\theta$. Accordingly, by defining the set:
\beq
{\cal S}_\ell=
\{(\eta,u)\in\mathbb{R}^2:
0\leq \eta\leq \theta,\,\, v_1\leq |u|\leq v_2
\},
\eeq
we have:
\beq
\int_{\frac{\delta}{\sqrt{\mu}}<|u|<\frac{a}{\sqrt{\mu}}}
|\tilde\varphi_{k,\mu}(u)|du
\leq
2\,\frac{a-\delta}{\sqrt{\mu}}\,c^{1/\mu+1},
\label{eq:finalinequality}
\eeq
where
\beqa
c&=&\prod_{\ell=1}^S\left(\max_{(\eta,u)\in{\cal S}_\ell} \frac{|\nu(\eta+j u)|}{\nu(\eta)}\right)\nonumber\\
&=&\prod_{\ell=1}^S \frac{|\nu(\bar \eta_\ell+j \bar u_\ell)|}{\nu(\bar \eta_\ell)}.
\label{eq:maximaCF}
\eeqa
In the above equation, $(\bar \eta_\ell,\bar u_\ell)$ are the pairs where the maxima are attained, which belong to ${\cal S}_\ell$ since 
$\frac{|\nu(\eta+j u)|}{\nu(\eta)}$, regarded as a function of the pair $(\eta,u)$,  is continuous.

In view of property P3, $\frac{\nu(\eta+j u)}{\nu(\eta)}$ is (as a function of $u$) nothing but the CF of $\bx_k(n)$ under an exponential measure translation. If $\bx_k(n)$ is not lattice~\cite{FellerBookV2}, the magnitude of the CF attains the value one only at the origin. This implies that all the maxima appearing in~(\ref{eq:maximaCF}) are strictly less than $1$, because $u$ is bounded away from zero in ${\cal S}_\ell$.  
Since, for any $c<1$, the function $c^{1/\mu}$ vanishes faster than any power of $\mu$ as $\mu\rightarrow 0$, we conclude that the integral in~(\ref{eq:finalinequality}) vanishes as $\mu$ goes to zero, and the proof of the lemma is complete.

~\hfill$\square$ 

\vspace*{10pt}
\noindent
\textsc{Remark VI}. 
The proof of Lemma 3 assumes that the local statistic $\bx_k(n)$ is not lattice. The distribution of a lattice random variable with span $\Delta$ is concentrated at the points $d,d\pm\Delta,d\pm 2\Delta,\dots$ for a certain real number $d$. The pertinent CF becomes:
\beq
\frac{\nu(\eta+ju)}{\nu(\eta)}=\frac{\displaystyle{\sum_{m=-\infty}^\infty} p_m e^{\eta (d+m\Delta)} e^{j t (d+m\Delta)}}{\nu(\eta)},
\eeq
where $p_m=\P[\bx_k(n)=d+m\Delta]$. Accordingly, we have
\beq
\frac{|\nu(\eta+jt)|}{\nu(\eta)}=\frac{\displaystyle{\left|\sum_{m=-\infty}^\infty p_m e^{\eta (d+m\Delta)} e^{j t m\Delta}\right|}}{\nu(\eta)},
\eeq
which shows that the magnitude of the CF has period $2\pi/\Delta$ and, in particular, it assumes the value one at all points $t=2\pi h/\Delta$, for $h=0,\pm1,\pm2,\dots$, which would violate the last step in our proof above. 

~\hfill$\square$ 



\end{document}